\documentclass {iopart}
\usepackage{iopams}

\usepackage{cite}
\usepackage{graphicx}
\usepackage{dcolumn}
\usepackage{bm}
\usepackage{xcolor}
\usepackage{import}
\usepackage{verbatim}
\usepackage{tikz}
\usetikzlibrary{arrows.meta}
\tikzset{%
  >={Latex[width=2mm,length=2mm]},
            base/.style = {rectangle, rounded corners, draw=black,
                           minimum width=5cm, minimum height=1cm,
                           text centered, font=\sffamily},
  activityStarts/.style = {base, fill=white},
       startstop/.style = {base, fill=white},
    activityRuns/.style = {base, fill=white},
         process/.style = {base, minimum width=3cm, fill=white, font=\ttfamily},
}
\usepackage{adjustbox}

\usepackage[normalem]{ulem}

\begin{document}

  \title{The effect of obstacles near a silo outlet on the discharge of soft spheres}

    \author{{Jing Wang$^{1}$}, {Kirsten Harth}$^{1,2}$, {Dmitry Puzyrev$^1$}, and { Ralf Stannarius$^1$} }

    \address{$^1$ Institute of Physics, Otto von Guericke University Magdeburg, Universit\"atsplatz 2, D-39106 Magdeburg, Germany \\
        $^2$ Department of Engineering, Brandenburg University of Applied Sciences, Magdeburger Stra\ss{}e 50, D-14770 Brandenburg an der Havel, Germany.}

\ead{jing.wang@ovgu.de}
\date{\today}

\date{Received: date / Accepted: date}
    \date{\today}

 \begin{abstract}
Soft smooth particles in silo discharge show peculiar characteristics, including, for example, non-permanent clogging and intermittent flow. This paper describes a study of soft, low-frictional hydro\-gel spheres in a quasi-2D silo. We enforce a more competitive behavior of these spheres during their discharge by placing an obstacle in front of the outlet of the silo. High-speed optical imaging is used to capture the process of discharge. All particles in the field of view are identified and tracked by means of machine learning software using a MASK R-CNN algorithm. With particle tracking velocimetry (PTV), the fields of velocity, egress time, packing fraction, and kinetic stress are analysed in this study. In pedestrian dynamics, it is known that the placement of an obstacle in front of a narrow gate may reduce the stress near the exit and enable a more efficient egress. The effect is opposite for our soft grains. Placing an obstacle above the orifice always led to a reduction of the flow rates, in some cases even to increased clogging probabilities.
    \end{abstract}

    \maketitle

\section{Introduction}

\label{sec:intro}

The storage of granular materials in silos and their discharge through small orifices in the container bottom has been studied widely since more than one century. The discharge dynamics of hard 
particles is practically independent of the pressure in the granular bed \cite{Perge2012pressure,Peralta2017pressure,Pongo2021hourglass}.
The outflow rate $Q$ of spheres is well described by Beverloo's 
equation \cite{beverloo1961flow,sperl2006experiments}. In two dimensions, it reads
\begin{equation}
Q = C \mu \sqrt{g}(W-kD)^{1.5},
\label{eq:one}
\end{equation}
where $\mu$ is the mass of the granulate per area, $D$ 
is the particle diameter, $W$ is the orifice width, $C$ and $k$ are dimensionless fitting constants. 
Hard particles can form permanent clogs when the outlet of a silo is smaller than a certain size \cite{thomas2016intermittency,alonso2021beverloo},  which for spheres is approximately five times the sphere diameter. 
The state of permanent clogging can only be destroyed by 
external force, such as vibration of the container or injection of pressurized air at the outlet. There are plenty of 
studies focusing on reducing the probability of clogging, such as 
expanding the orifice size \cite{janda2009flow}, adjusting the number of outlets and distances between outlets \cite{mondal2014role,kunte2014spontaneous,xu2018inter}, or applying external vibration \cite{chen2006flux,wen2015flux}.
Another effective method in some situations is to place an obstacle inside the silo. The presence of such an obstacle in a certain height above the outlet 
may decrease the probability of clogging. The mechanics behind this effect is assumed to be the reduction of pressure in the 
region of arch formation \cite{zuriguel2011silo,lozano2012flow}.  
The packing fraction at the orifice decreases when putting an obstacle in front of the outlet \cite{arean2020granular}.  
When the obstacle is placed in an optimal position in front of the 
orifice, the outflow rate can even increase \cite{alonso2012bottlenecks}. These observations were made with hard, frictional grains. However, other related systems such as pedestrians or animals exiting through a door display similar features.

Pedestrian motion in narrow passages at high population density shows some similarities with 
granular dynamics \cite{helbing1998computer}. Although all
pedestrians have individual trajectories and destinations, their motion is principally 
influenced by repulsive interactions with other individuals at high population 
densities. This can lead to self-organized dynamics\cite{helbing2001self}. At very high density, pedestrians can
organize in lanes \cite{oeding1963verkehrsbelastung}, which is similar to 
segregation in granular media \cite{makse1997spontaneous}. The behavior of pedestrians passing bottlenecks such as corridors or doors is in some respect analogous to granular flow in an hourglass \cite{wieghardt1975experiments}. Density is used as a key parameter, closely related to pedestrian dynamics \cite{wolf1996traffic}, 
but kinetic stress, i.e. the product of packing density and velocity fluctuations, is considered as another key factor leading to velocity changes towards the 
bottleneck \cite{garcimartin2017pedestrian}. If pedestrians stay in a confined, not too 
wide space, the density at a bottleneck will surprisingly reduce \cite{adrian2020crowds}. If there is an obstacle in front of the door, it can help alleviate the pressure in 
the area near the door and decrease the probability of blocking the passage \cite{zhao2017optimal,zuriguel2020contact,echeverria2020pedestrian,Gella2022obstacle}. However, an obstacle does not necessarily alter the pedestrian flow rate \cite{garcimartin2018redefining}. Some studies point out the 'faster-is-slower' effect \cite{garcimartin2014experimental,pastor2015experimental}, which means 
that pedestrians would pass the bottleneck faster if they are under lower stress. A relation to friction was proposed~\cite{garcimartin2017pedestrian}. In a simulation, an ordered or 'crystalline' structure was helpful for faster egress dynamics \cite{cheng2021ordering}.

Soft granular materials, as studied here, are typically found in agriculture and pharmacy. They combine properties of hard granular matter with some features of animate objects. Soft grain ensembles exhibit quite different characteristics in silo discharge as compared to hard particles and show qualitatively novel features  \cite{ashour2017silo,hong2017clogging,desmond2013experimental,stannarius2019packing,harth2020intermittent,Tao2021soft} regarding clogging, intermittency of flow and avalanche statistics. This refers in particular to hydrogel spheres, as elastic granular material with very low friction that is incompressible but can easily be deformed. Elastic moduli are of the order of 10 to 100 kPa \cite{chippada2010simultaneous}. In silo discharge, 
the outflow rate depends on the pressure at the bottom of the container and thus on the fill height  \cite{ashour2017silo}. These soft particles can spontaneously rearrange themselves during transient congestions \cite{harth2020intermittent}.
We note that cattle or
human egress can show similar intermittent dynamics when passing a bottleneck \cite{Zuriguel2014sheep}. 
In the present study, we explore the effects of an obstacle placed in front of a container outlet for soft, low-frictional hydrogel spheres. 
An optical camera is used to record the process of discharge 
focusing on the region near the obstacle. Machine-learning based software, namely, a Mask Region-Based Convolutional Neural Network (MASK R-CNN) was employed to precisely locate each 
particle. With the coordinates of the 
particles, and a high enough video frame rate, it is possible to track each particle in successive frames. 
In addition to measurements of the actual discharge rates as a function of the obstacle position, we also extract flow fields, local packing fractions and the distribution of kinetic stress in the vicinity of the outlet. The latter is related to the pressure during the discharge.

\section{Experimental setup}
\label{sec:Setup}
\begin{figure*}[htbp]
\includegraphics[width=0.95\textwidth]{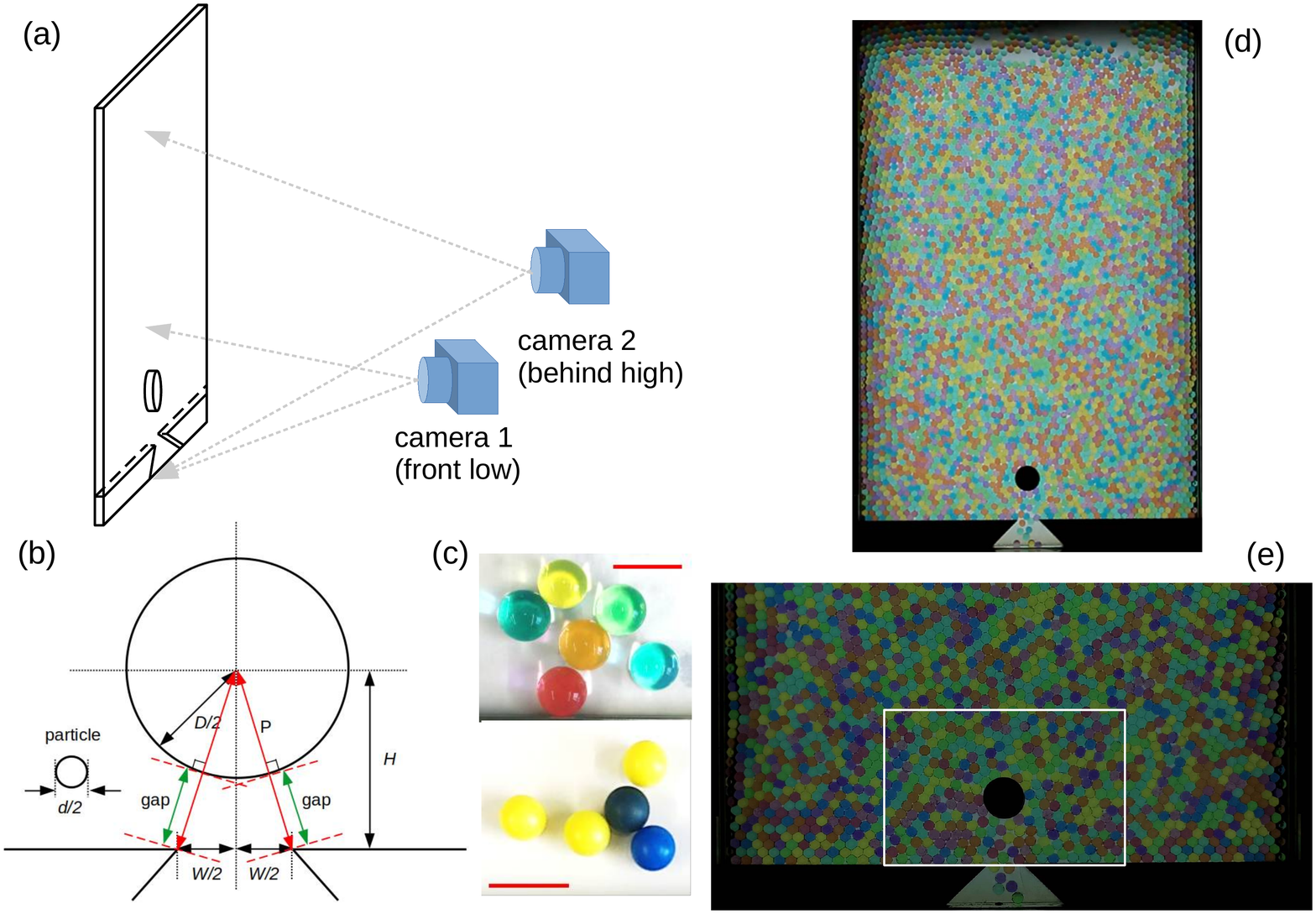}
\caption{\label{fig:sketch}(a)  Sketch of the experimental setup. Two cameras in front of the setup capture the discharge process. Camera 2, sketched on the right, monitors the height change during the discharge with 60 fps, (d) show one typical frame. Camera 1 is placed closer to the container and at lower height to record a detailed view of the region near the outlet at  and (e) is a typical frame recorded with the faster camera 1. The white rectangle sketches the $768~\times~512$ pixel$^2$ area of interest, cropped from the original $1920\times1080$ pixel$^2$ frame. (b) sketches the geometry near the obstacle above the orifice. (c) shows the particles, the top ones are hydrogel spheres (HGS) with 7 mm diameter, below are the airsoft bullets (ASB) with 6.5 mm diameter. Scale bars are 10 mm. }
\end{figure*}

The setup consists of a flat box of 80~cm height and 40~cm width. The depth 
can be adjusted to the particle diameter to ensure 
that the container holds only one vertical layer of particles. The orifice 
is formed by the adjustable gap between two aluminum 
sliders at the container bottom \cite{ashour2017silo}. A sketch of the setup is shown in 
Fig. \ref{fig:sketch}. 
Two 3~cm wide aluminum bars holding the front and back glass plates hide the edges, and the observable area in the pictures is 34~cm. The bin is much higher than the height of the 
observable area. In the present study, the bin is filled with particles up to a filling height of approximately 60~cm before each experimental run, which is slightly higher than the observable area.

The obstacle is realized by a cylindrical magnet of 25~mm diameter and 3~mm thickness that is suspended inside the bin. It is held in place by a second magnet outside the bin. By this fixation method, we avoid suspension wires inside the silo but still keep the opportunity to change the obstacle position straightforwardly. The magnets are set above the center of the 
orifice. The height is defined as the position of the center of the cylinder above the orifice, labeled in Fig.~\ref{fig:sketch} by ${H}$.

Figure~\ref{fig:sketch}(b) details the geometry near the obstacle and the orifice in the silo. $D$ is the diameter of the circular obstacle, $d$ is the diameter of the spherical particles, $W$ is the width of the orifice in the silo, $H$ is the height position of the obstacle, viz. the distance of its center from the orifice level.

The width of the gaps between the obstacle and the edges of the outlet is given by
\begin{equation}
H' = \rho'd = \sqrt{H^2 + {\left(\frac{W}{2}\right)}^2}-\frac{D}{2}.
\label{eq:seven}
\end{equation}
The silo discharge is observed by two commercial video cameras, see Fig.~\ref{fig:sketch}(a). The front 
 camera (XiaoYi 4K+ Action) focuses on the 
bottom area around the obstacle and captures videos with a 
frame rate of 120 fps with a spatial resolution of 0.2~mm per 
pixel. The second camera (Canon EOS 600D) is placed behind and above the front one and captures videos with a frame rate of 60 fps to monitor the height changing in the silo during discharge.

Commercial hydrogel spheres (HGS) are swelled for at least 24 hours in NaCl 
solution. The size of the 
swollen HGS depends on the salt concentration. We select a NaCl
concentration of 5.0 g/L and obtain HGS with a mean
diameter of $d=7$~mm, which varies by approximately 3~\%. 
An elastic modulus of roughly 30~kPa was determined using the 
Hertzian contacting model \cite{mindlin1953elastic} by measuring the diameter of the contact area under given weights.

In order to compare the differences of the behavior of soft, 
frictionless grains to hard, frictional particles of a similar 
size and fraction, we have conducted additional experiments with hard
plastic spheres (airsoft bullets, ASB) of $d=6$~mm diameter and a friction coefficient of approximately 0.3. They behave as rigid spheres.
The experimental parameters are listed in Tab. \ref{tab:table1} and the geometry is sketched in Fig.~\ref{fig:sketch}(b), where $\rho$ is the ratio of the orifice width to the particle diameter, and $\rho '$ is the ratio of the gap width to the particle diameter.
\begin{table}[b]
\caption{\label{tab:table1} Experimental conditions: Materials, orifice widths $W$, ratio $\rho=W/d$, ratio $\rho'=W'/d$  and obstacle position $H$\bigskip}

\begin{tabular}{|lcccr|}
\hline\textrm{Material}&
\textrm{Orifice width $W$}&
\textrm{$\rho$}&
\textrm{$\rho '$}&
\textrm{Height $H$}\\
\hline
HGS & 14 mm & 2.0 & 0.9 & 17.5 mm\\
HGS & 14 mm & 2.0 & 2.3 & 27.5 mm\\
HGS & 14 mm & 2.0 & 3.7 & 37.5 mm\\
HGS & 14 mm & 2.0 & - & -\\
HGS & 21 mm & 3.0 & 1.1 & 17.5 mm\\
HGS & 21 mm & 3.0 & 2.4 & 27.5 mm\\
HGS & 21 mm & 3.0 & 3.8 & 37.5 mm\\
HGS & 21 mm & 3.0 & - & -\\
\hline
ASB & 35 mm & 5.8 & 4.8 & 37.5 mm\\
ASB & 35 mm & 5.8 & - & -\\
\hline
\end{tabular}

\end{table}

\section{Methods}
\label{sec:Methods}

The optical images are evaluated sequentially to determine particle positions, local packing fractions, velocities and other dynamic features of the material in the silo, with particular emphasis on the regions near the outlet and the obstacle. The necessary evaluation steps are described in detail in the appendix. Data processing starts with a machine-learning aided identification of all particles in the field of view and the determination of their positions. This is done with a MASK R-CNN algorithm as described in \ref{app:Mask-rcnn}. From the tracking of individual particles in successive images, their velocities are derived to construct a velocity field in the container (\ref{app:velocities}). An interesting quantity, in particular for egress dynamics, is the escape time from a certain starting position in the silo, which is derived from particle tracking as described in \ref{app:escape}. The positions of individual grains are averaged over a certain number of frames to obtain an average packing local fraction (\ref{app:packing}). The kinetic stress can be determined from velocity fluctuations as described in \ref{app:kineticstress}.
The term of kinetic stress in DEM models \cite{gollin2017extended,weinhart2016influence} is analogous to the pressure on particles induced by fluctuations. 
It results from the correlation between the local packing fraction, which can be considered as degree of deformation, and fluctuations of the velocity.

\section{Results}
\label{sec:Results}

\begin{figure}[htbp]
\centering
\includegraphics[width=0.75\columnwidth]{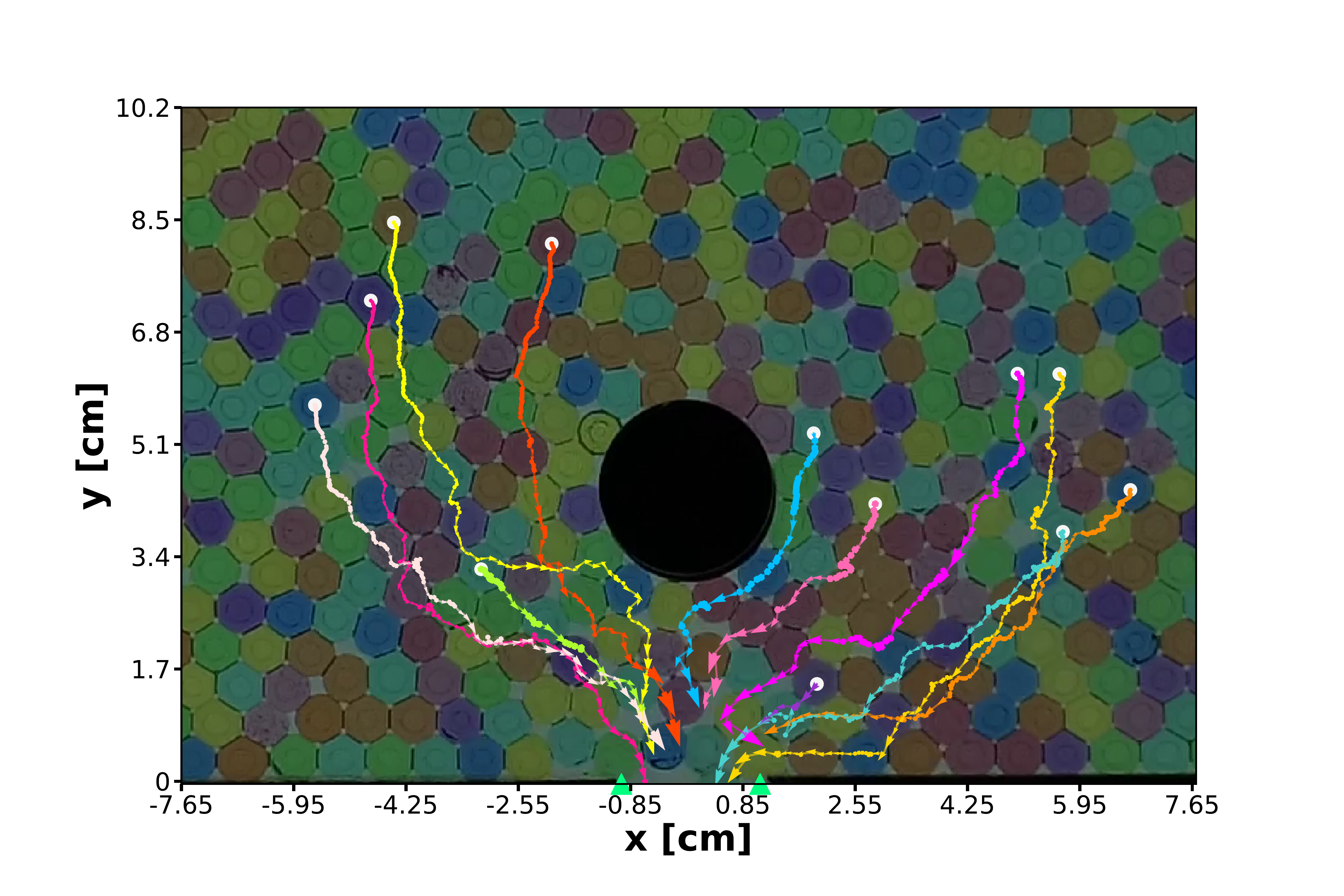}
\caption{\label{fig:traj}Trajectories of selected HGS particles.}
\end{figure}

Figure \ref{fig:traj} shows typical trajectories of 12 selected particles. White markers correspond to the initial positions of these particles. The colored lines are their paths towards the outlet of the silo. Small arrows of corresponding color along the trajectories symbolize the instant velocities.

\begin{figure*}[htbp]
\includegraphics[width=1.0\textwidth]{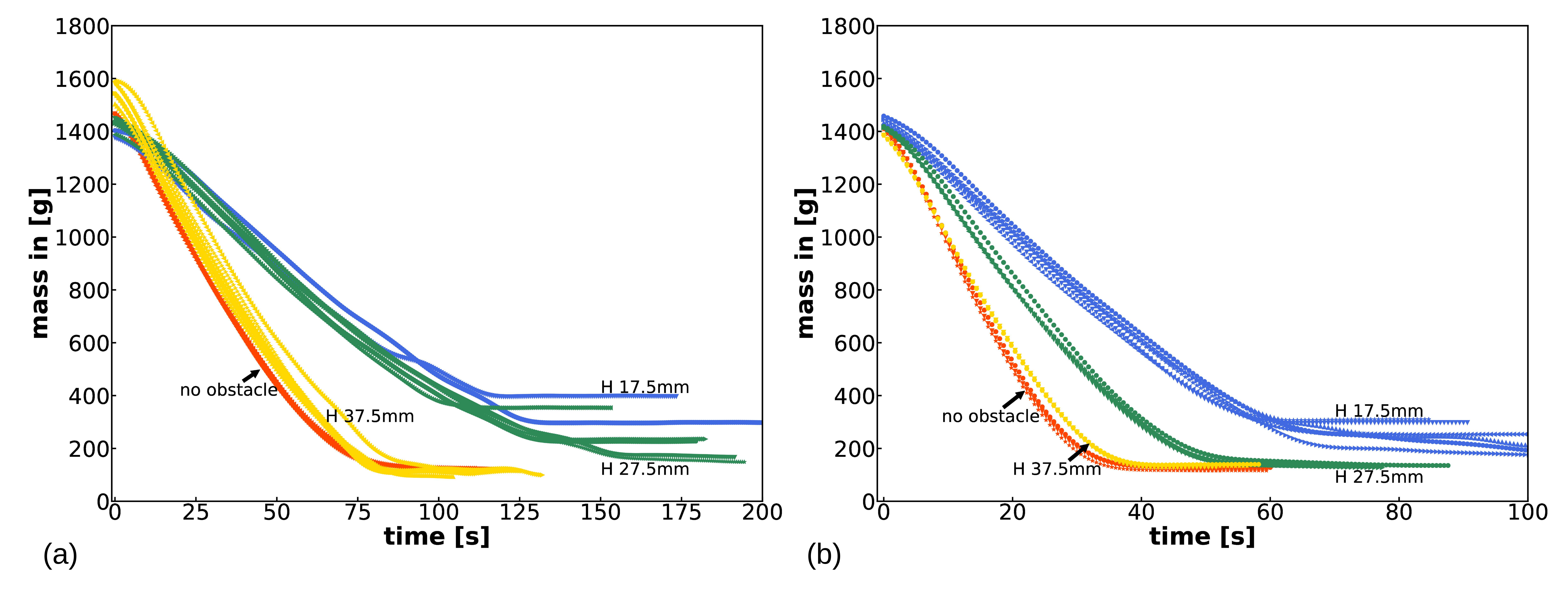}
\caption{\label{fig:mass2} (a) HGS Mass remaining in the silo with (a) $W=14$~mm and (b) $W=21$~mm orifice widths. Some graphs reach plateaus when the silo discharge stops before it is emptied completely. Initially, the silos are filled to a height of approximately 60~cm.}
\end{figure*}

Before the structure and dynamics of the flow field inside the silo is analyzed in detail, the main consequences of an obstacle on the global outflow dynamics are shown in Fig.~\ref{fig:mass2}. 
It compares the mass discharge rates of silos with two different orifice widths $W$ and different obstacle positions. All silos started with approximately the same initial mass of about 1.5~kg (60~cm fill height). For each obstacle geometry, several independent runs are shown. During the initial phase of the discharge, there is only little difference between the individual runs for identical obstacle and orifice geometries, but it is clearly seen that the placement of an obstacle in front of the outlet has significant consequences when it is placed closer to the outlet. In fact, the observed scenarios are remarkable. There are, in principle, two effects that one has to consider: First, when the gap width $\rho'$ becomes small enough, it forms a bottleneck for the discharge. Second, for the soft grains, a lower pressure near the outlet retards the discharge and leads to clogging at the outlet \cite{ashour2017silo}. When the obstacle is close enough, directly above the opening, the pressure below is reduced and the particle behave more rigidly, thus discharging more slowly and clogging even when the outlet is larger enough for a complete discharge without obstacle. The silo does not empty completely anymore. This is indicated in Fig.~\ref{fig:mass2} by the different plateau levels reached in the graphs. They indicate that material remains in the silo and congestions form. 100~g of material correspond to an average fill height of 4 cm. Note that we did not prove whether these plateaus are permanent congestions or whether they dissolve on a time scale of several minutes, the experiments were stopped after clogs lasted for more than 30 seconds.

\begin{figure}[htbp]\centering
\includegraphics[width=0.5\columnwidth]{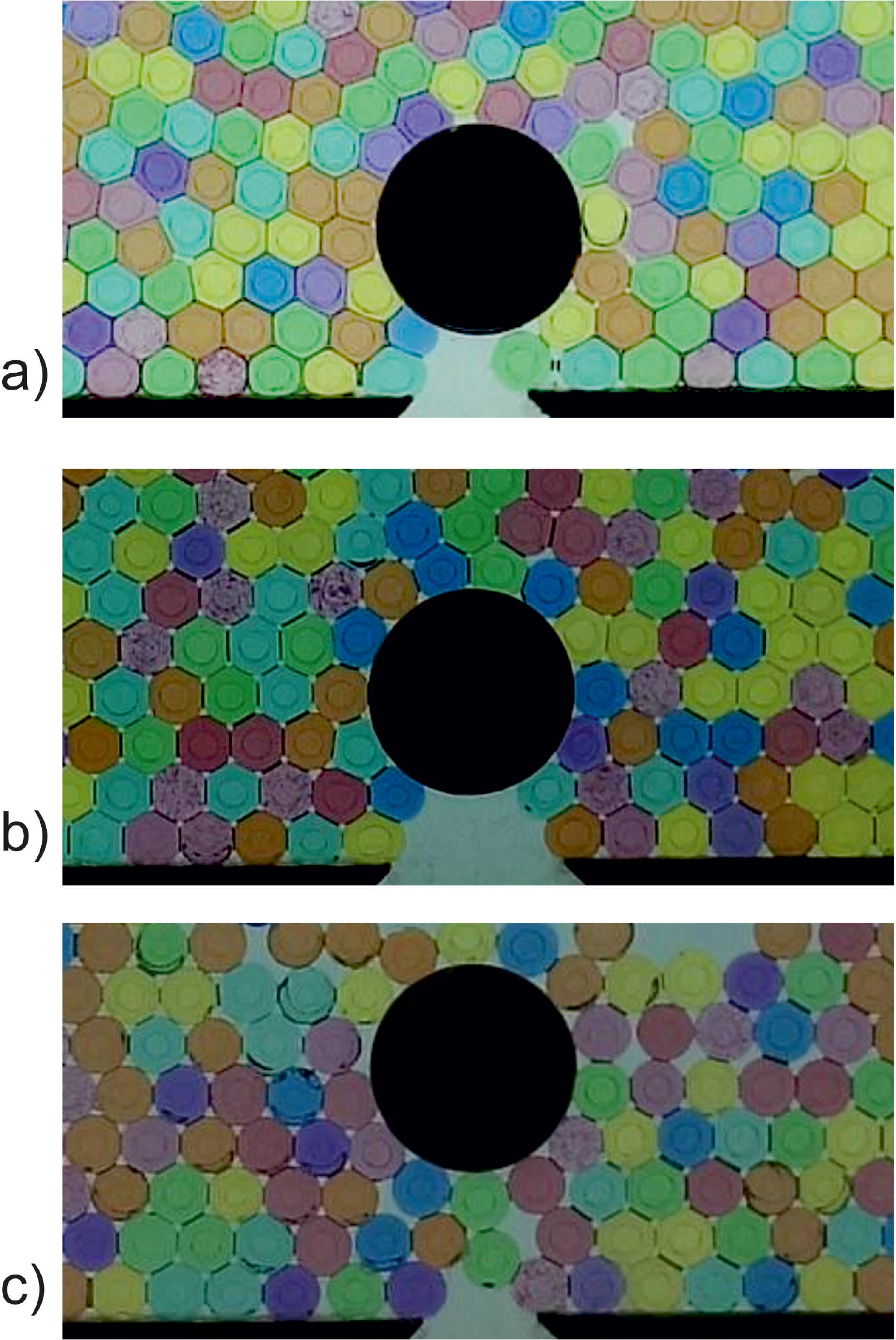}
\caption{\label{fig:clogs} Long lasting (or possibly permanent) clogged states of HGS in presence of an obstacle above the orifice, (a) $W=14$~mm ($\rho=2$), $H=17.5$~mm ($\rho'=0.9$), (b) $W=21$~mm ($\rho=3$), $H=17.7$~mm ($\rho'=1.1$), and (c) $W=14$~mm ($\rho=2$), $H=27.5$~mm ($\rho'=2.3$).}
\end{figure}

It is even more interesting to compare the different situations and to have a more quantitative view. This reveals that there are two different scenarios depending on the obstacle position. At the lowest position, where $\rho'$ is of the order of one (gap width comparable to, and even lower than the original particle diameter), the outflow still proceeds as long as the fill level is sufficiently high. The pressure squeezes the particles through the gaps even when $\rho'<1$. This is remarkable because when $\rho$ is well below 2 in a silo without obstacle and orifice at the bottom, clogging would set in much earlier. Obviously, the soft grains are less prone to clogging when the bottleneck is not horizontal. 
In presence of the obstacle near the outlet, when the silo is partially emptied and the pressure at the bottom drops, the particles clog at these lateral gaps, the area below the orifice is then free of material (see Fig.~\ref{fig:clogs} (a,b)). When the obstacle is placed further away from the outlet, and $\rho'$ is comparable to or larger than $\rho$, this does not happen, but we find the second clogging mechanism mentioned above (Fig.~\ref{fig:clogs}(c)). It can be seen from the nearly spherical appearance of the HGS below the obstacle that they are exposed to small forces only, in contrast to the particles at both sides.    

\subsection{Velocity}
\label{sec:velocity}
\begin{figure}[htbp]
\centering
\includegraphics[width=0.6\textwidth]{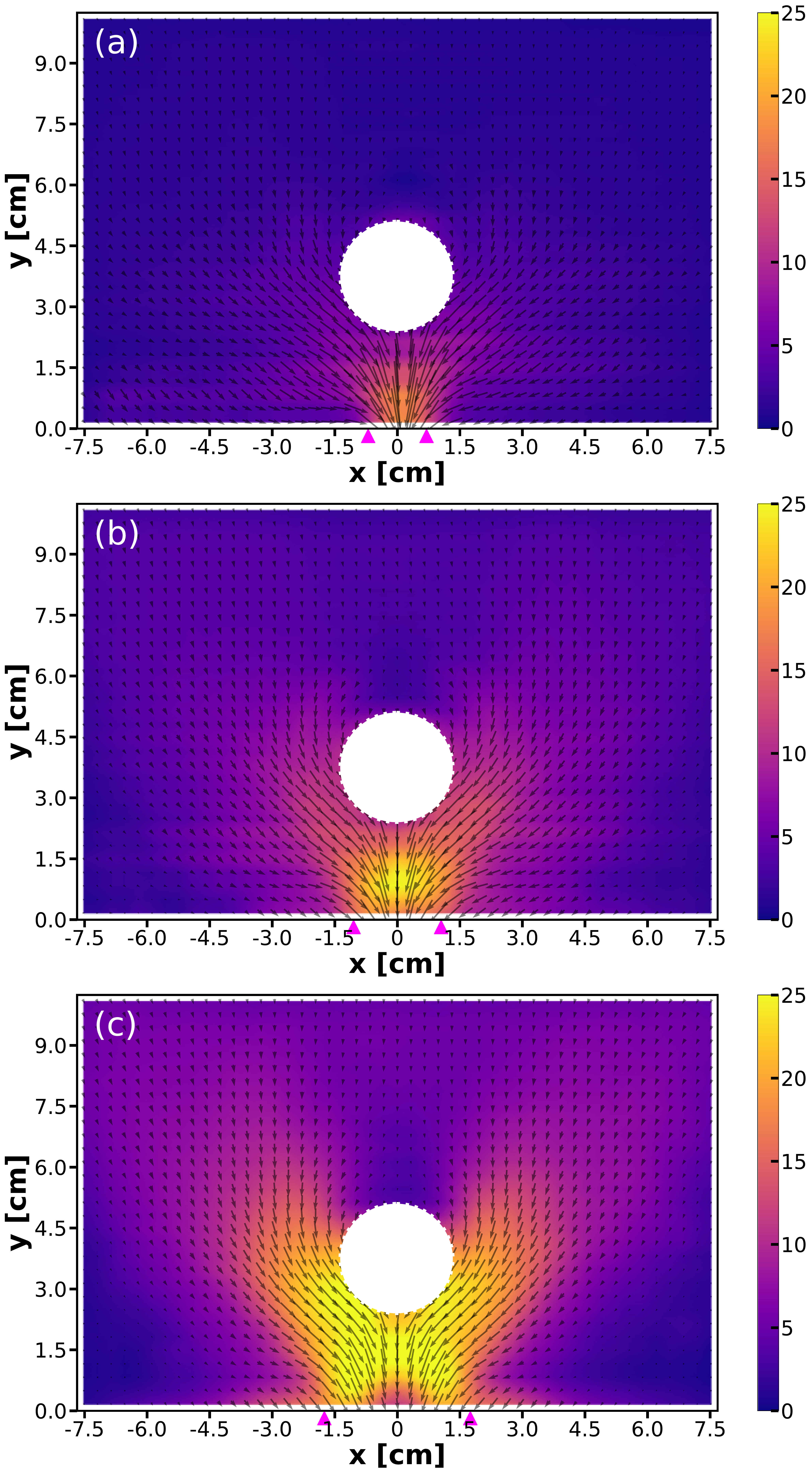}
\caption{\label{fig:vel_h5} Velocity fields in a silo with an obstacle at height $H=37.5$~mm. (a) HGS, orifice width $W=14$~mm, (b) HGS, $W= 21$~mm, (c) ASB $W= 35$~mm ($W\approx 5.8d$). The velocity fields are averaged over a time period where the silo empties from half to quarter filling. Color bars represent the velocity in units of cm/s. }
\end{figure}

Figures~\ref{fig:vel_h5}(a,b) show the velocity distribution for the soft HGS in the region near orifices of different widths, with the obstacle at a fixed height of $H=37.5$~mm ($\rho'\approx 3.75~d$). For comparison, Fig.~\ref{fig:vel_h5}(c) shows the velocity distribution of an ensemble of hard ASB, albeit with a much wider orifice width. 
There, one can observe distinct triangular areas at both bottom corners, where the hard frictional ASB particles hardly move. These are the edges of stagnant zones.
Immediately above the obstacle, particles slow down and almost come to rest.

It is not possible to directly compare the hard and soft sphere experiments because of the quite different orifice sizes. When the hard grains are observed with smaller outlet widths, there is no continuous flow because of frequent clogging, and when the soft grains are observed with larger orifice sizes, the packing fraction is no longer homogeneous. Empty holes are formed that propagate upwards while the particles are in free fall in these regions. The reason for the soft grains moving much slower next to the obstacle than the hard spheres is the smaller orifice width that forms a bottleneck for the outflow. Since the orifice width $W$ is much larger for the ASB, this bottleneck effect is much less dramatic there.

\begin{figure*}[htbp]
\includegraphics[width=1.0\textwidth]{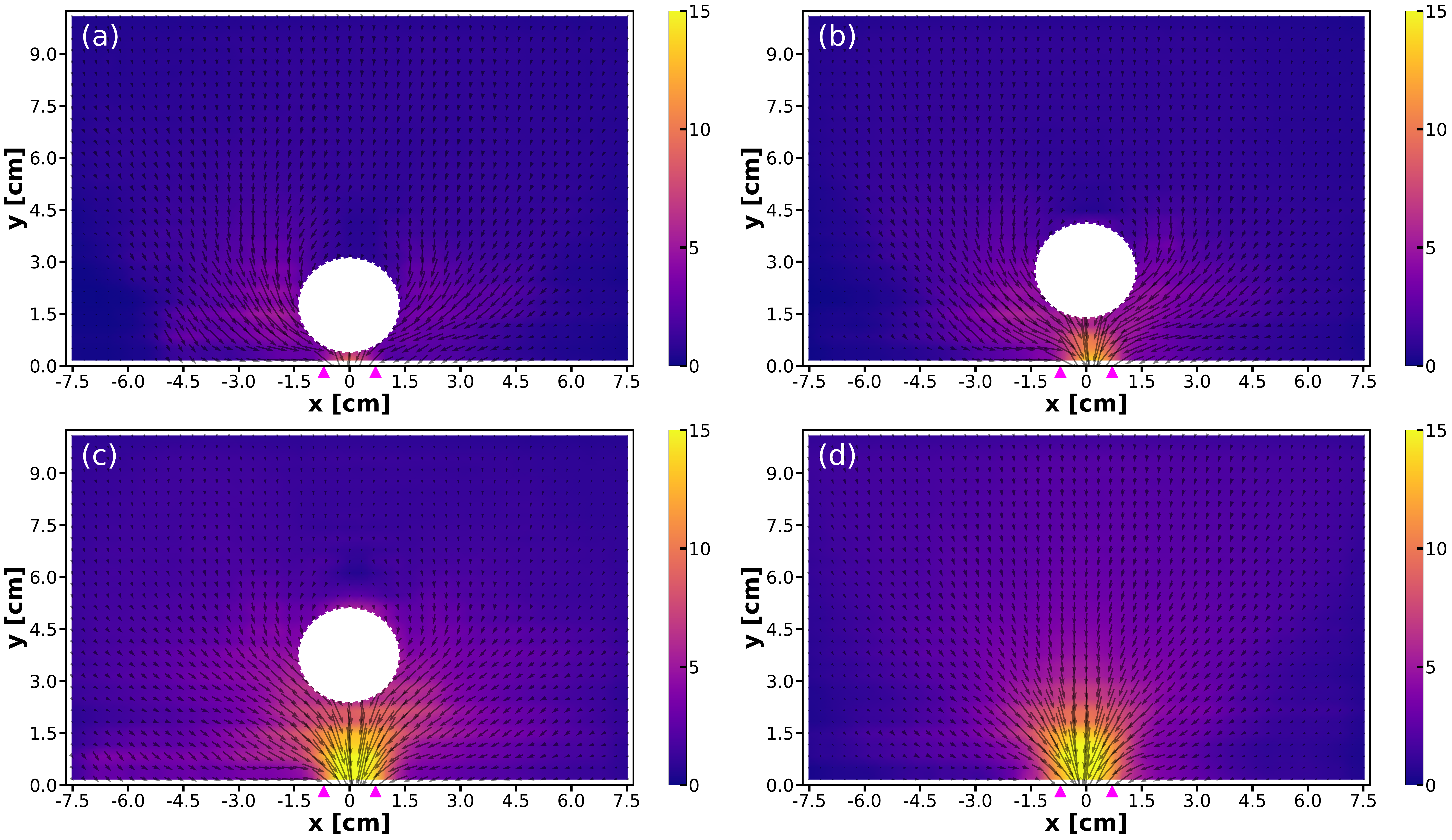}
\caption{\label{fig:vel_14} Velocity fields of HGS in the silo at $W=14$~mm orifice width with an obstacle at height of (a) $H=17.5$~mm, (b) $H= 27.5$~mm and (c) $H=37.5$~mm. (d) shows the velocity field of in absence of an obstacle. The velocities are time averaged over period where the silo discharges from half-filled to quarter-filled. Color bars indicate the velocities (in units of cm/s)}
\end{figure*}

Figure~\ref{fig:vel_14} shows the velocity distribution in the silo with $W=2d=14$~mm and an obstacle at different heights above the orifice. Compared with the case of no obstacle shown in Fig.~\ref{fig:vel_14}(d), the velocities near the orifice are systematically lower when an obstacle is present. 
When the obstacle height reaches  37.5~mm (Fig.~\ref{fig:vel_14}(c), $\rho'=3.7$)
or more, the velocity directly above the orifice has practically reached the same value as in absence of  obstacles.
However, a difference in the flow field pattern can be still observed. With the obstacle, the flow profile is broadened and extends more towards the side walls, which is intuitively clear.

\begin{figure*}[htbp]
 
\includegraphics[width=1.0\textwidth]{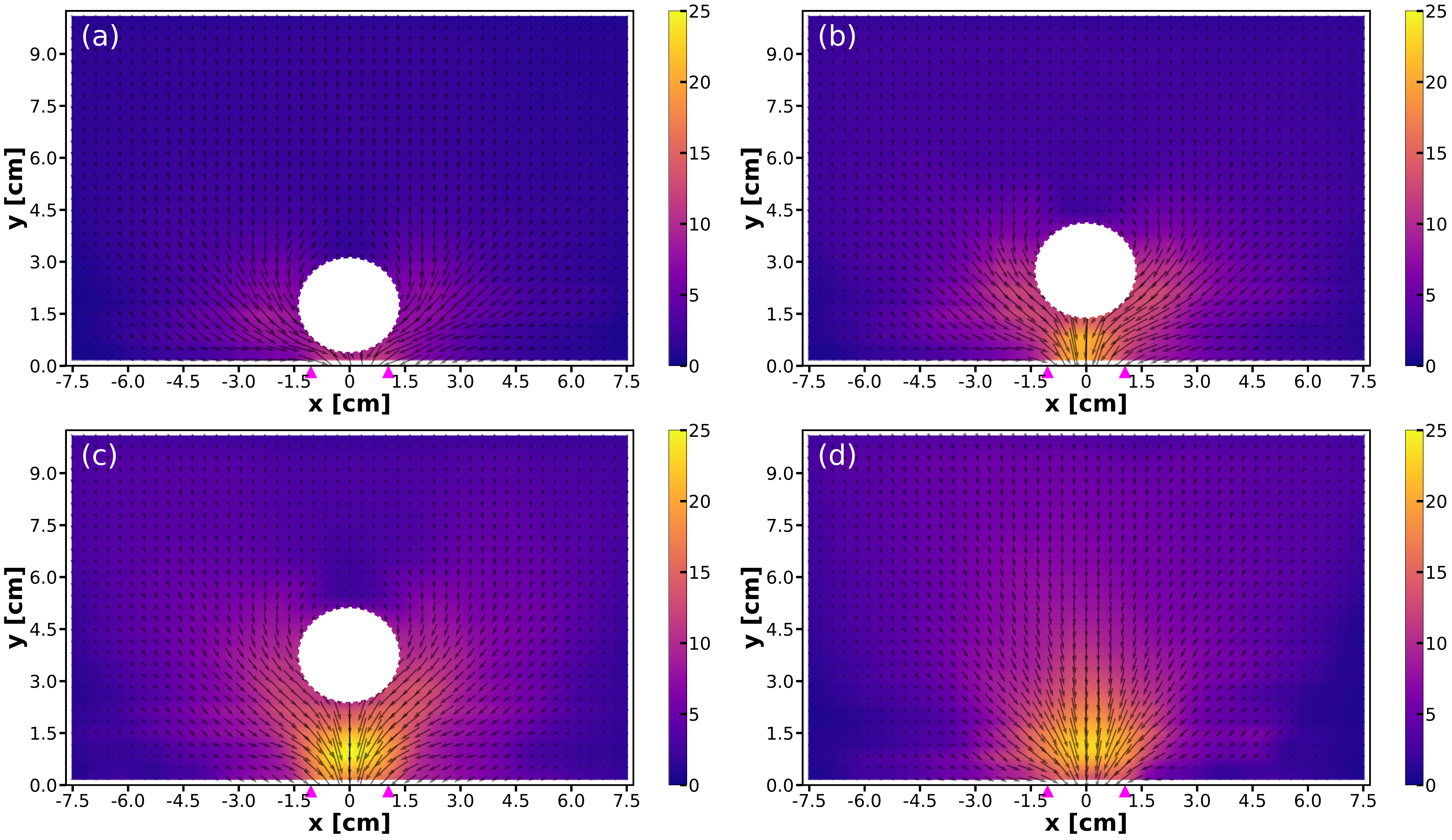}
\caption{\label{fig:vel_21}(a) Velocity fields of HGS at $W=21$~mm orifice and the obstacle at (a)  $H=17.5$~mm, (b) $H= 27.5$~mm, (c) $H= 37.5$~mm and (d) velocity field in absence of an obstacle. The velocity fields are time averaged over a period from the half-filled to the quarter-filled silo. The colors indicate the velocities (in units of cm/s)}
\end{figure*}

Figure~\ref{fig:vel_21} presents the velocity map in the silo with $W=21$~mm orifice width ($W=3d$) and the obstacle at different heights. 
In general, with the wider orifice (by one particle diameter with respect to Fig.~\ref{fig:vel_14}), particles can flow out more smoothly and the velocity maximum increases.
Even at the lowest obstacle height of 17.5~mm, the outflow through the larger orifice is faster because of the slightly larger gap width $\rho'$ (Figs.~\ref{fig:vel_14}(a),\ref{fig:vel_21}(a)).

\begin{figure}[htbp]
\centering
\includegraphics[width=0.75\columnwidth]{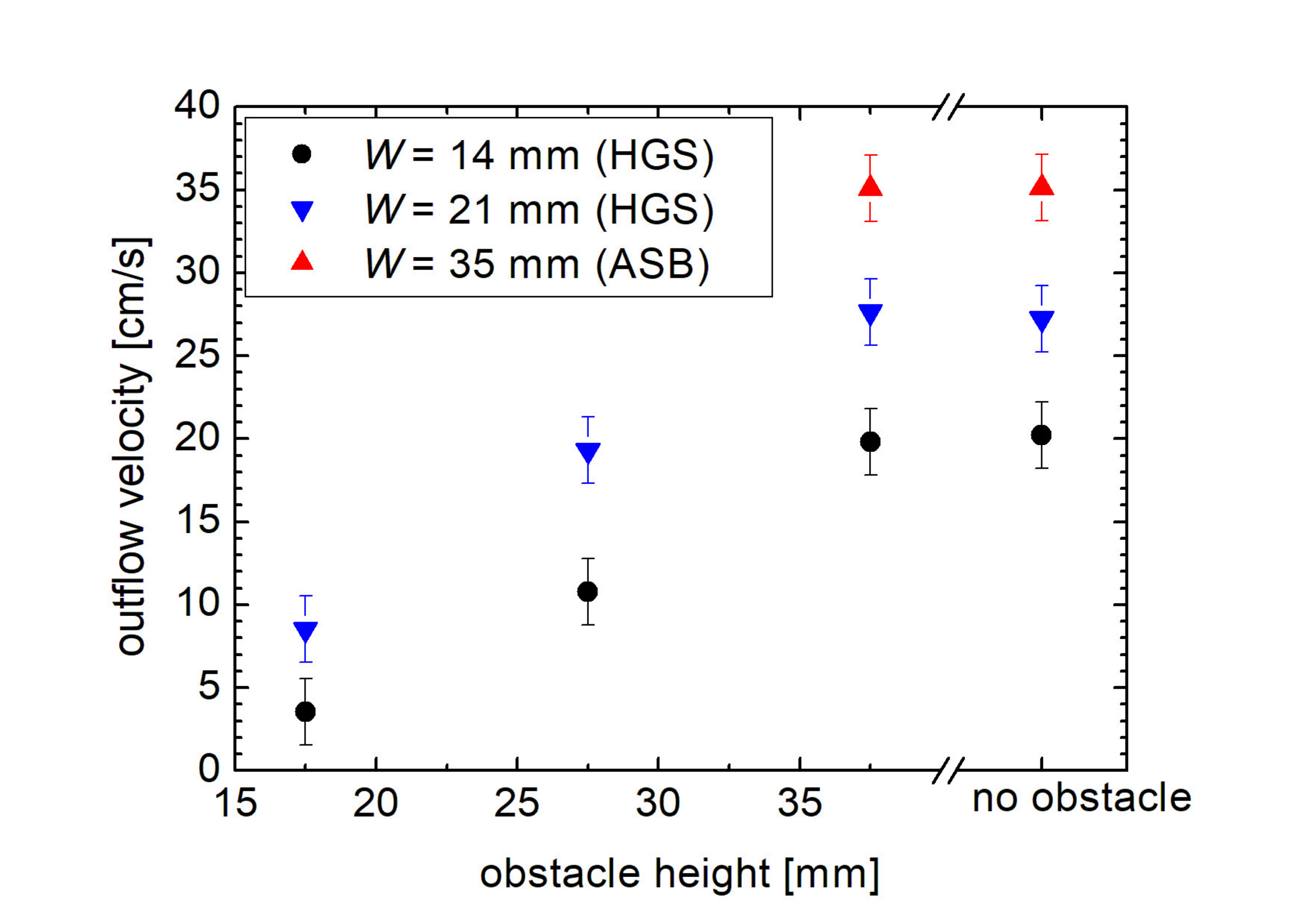}
\caption{\label{fig:vel_max} Time-averaged velocity of particles leaving the orifice in the center ($\pm 6$ mm). The rightmost symbols were measured in absence of an obstacle.}
\end{figure}

The mean vertical velocity of particles leaving the container is plotted in Fig.~\ref{fig:vel_max}. This graph demonstrates that within the experimental reproducibility, there is practically no difference in the outflow velocity when the obstacle is in the highest position ($\rho'\approx 3.7$) above the outlet.
For lower positions of the obstacle, the outflow velocity of the soft grains is affected significantly. This is particularly interesting for obstacle height 27.5~mm where the width of each of the two gaps is still larger than the outlet at the bottom. The primary cause for the reduced outflow velocity is a lower pressure on the spheres.

\clearpage

\subsection{Escape time}
\label{sec:Exittime}

\begin{figure}[htbp]
\centering
\includegraphics[width=0.6\textwidth]{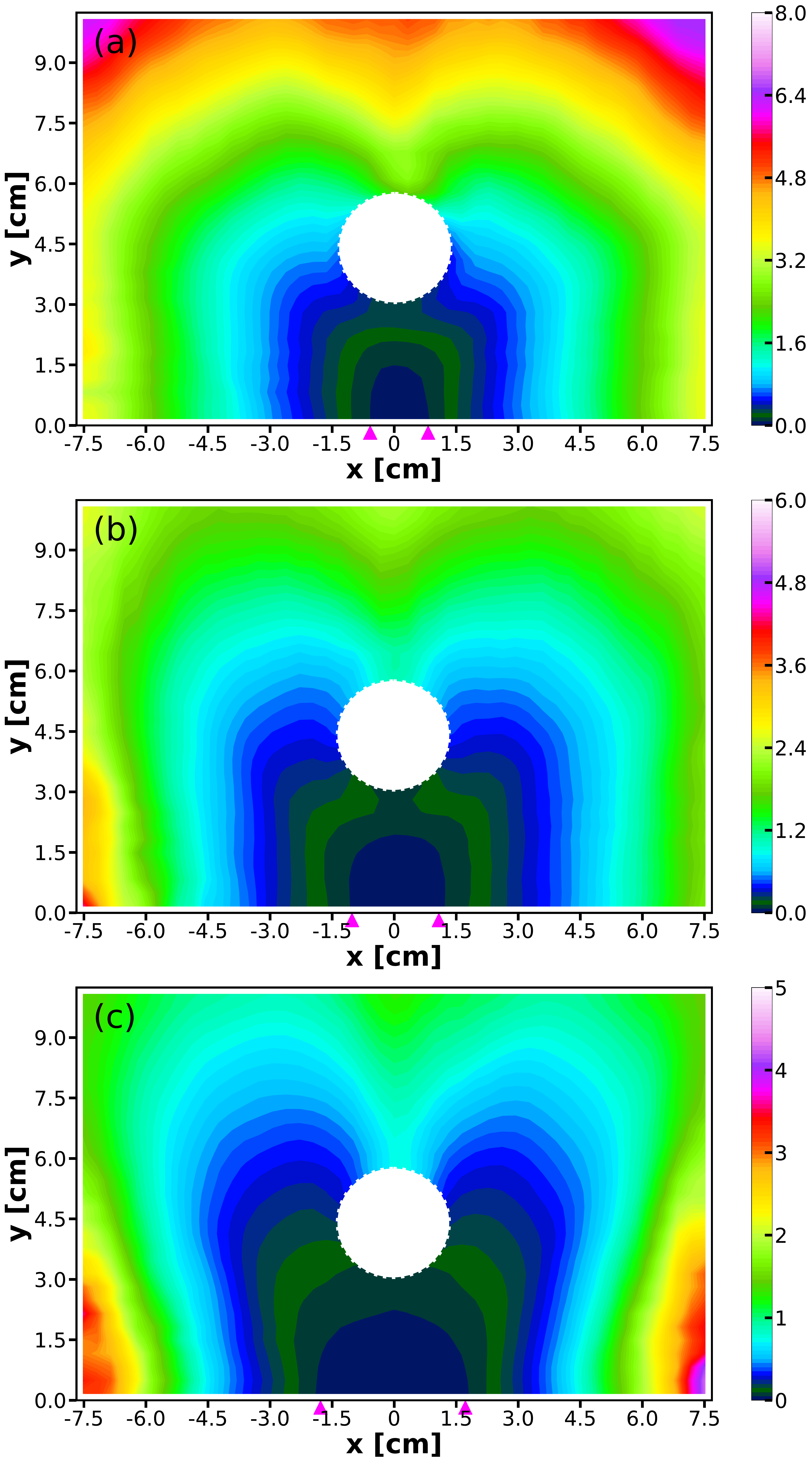}
\caption{\label{fig:esc_h5} Distribution of escape times in a silo with an obstacle at a height of 37.5~mm.
(a) HGS with $W=14$~mm, (b) HGS with $W=21$~mm, (c) ASB with $W=35$~mm. Escape times were averaged over the period where the silo emptied from half to quarter filling. The color bars give time in seconds.}
\end{figure}

The spatial distribution of escape times is of particular importance in human egress dynamics \cite{garcimartin2017kinetic}, but it can also be a
decisive parameter in silo discharge. Averaged escape times of particles in a given starting position are presented in Fig.~\ref{fig:esc_h5} for different orifice widths and materials at a fixed obstacle position.
Note that the selected regions do not show the full silo width but only approximately one half of it.
In Fig.~ \ref{fig:esc_h5}(c), the two bottom corners indicate the transition to the stagnant zones of hard particles. 

\begin{figure*}[htbp]
\includegraphics[width=1.0\textwidth]{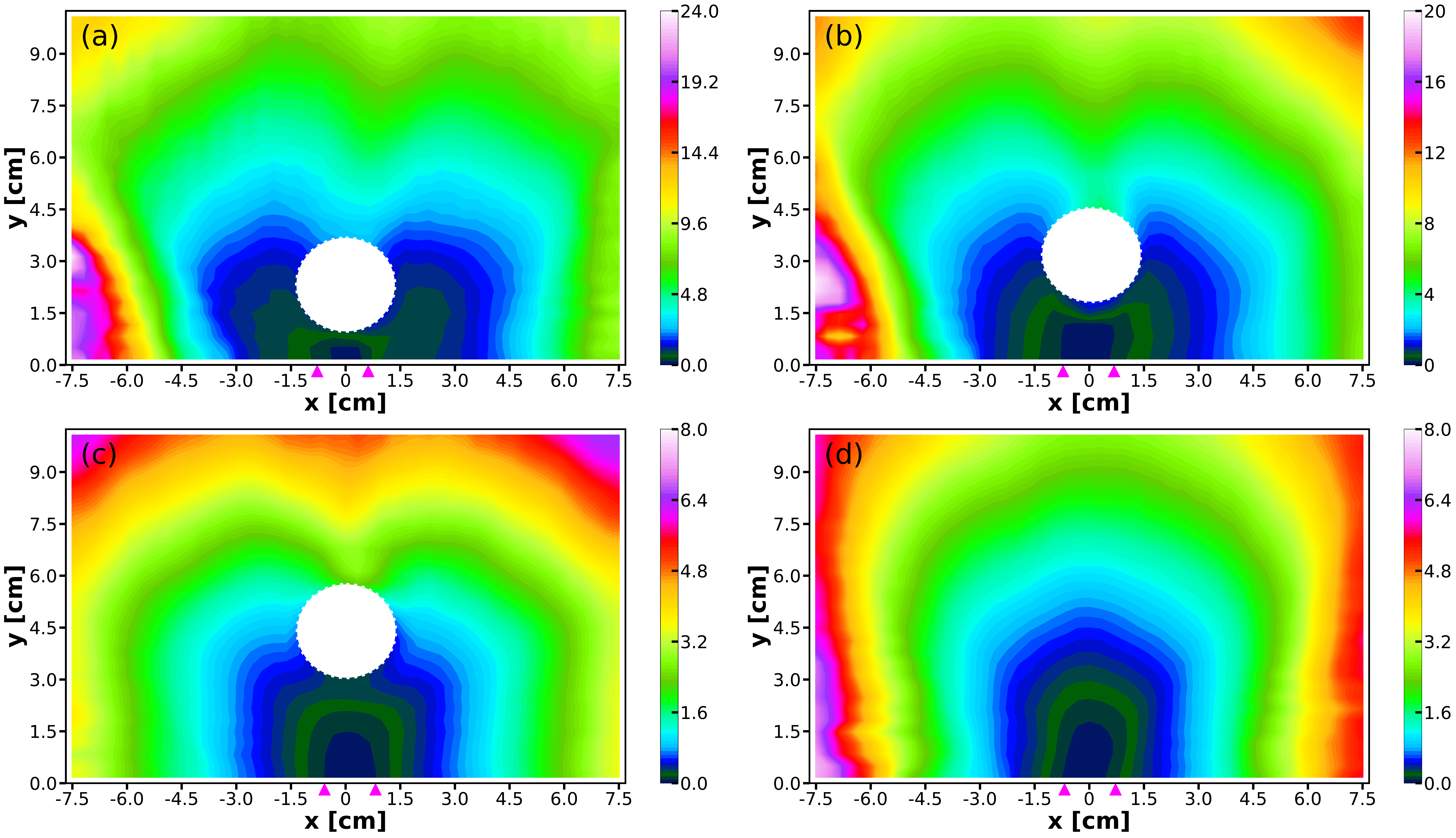}
\caption{\label{fig:esc_14}Distribution of escape times of HGS in the silo of $W=14$~mm orifice width with an obstacle at heights of (a)  $H=17.5$~mm, (b) 27.5~mm, and (c) 37.5~mm. (d) shows the escape times when the obstacle is absent. The escape times are averaged over the period where the silo was emptied from one half to one quarter. The color bar gives the time in seconds.
}
\end{figure*}

\begin{figure*}[htbp]
\includegraphics[width=1.0\textwidth]{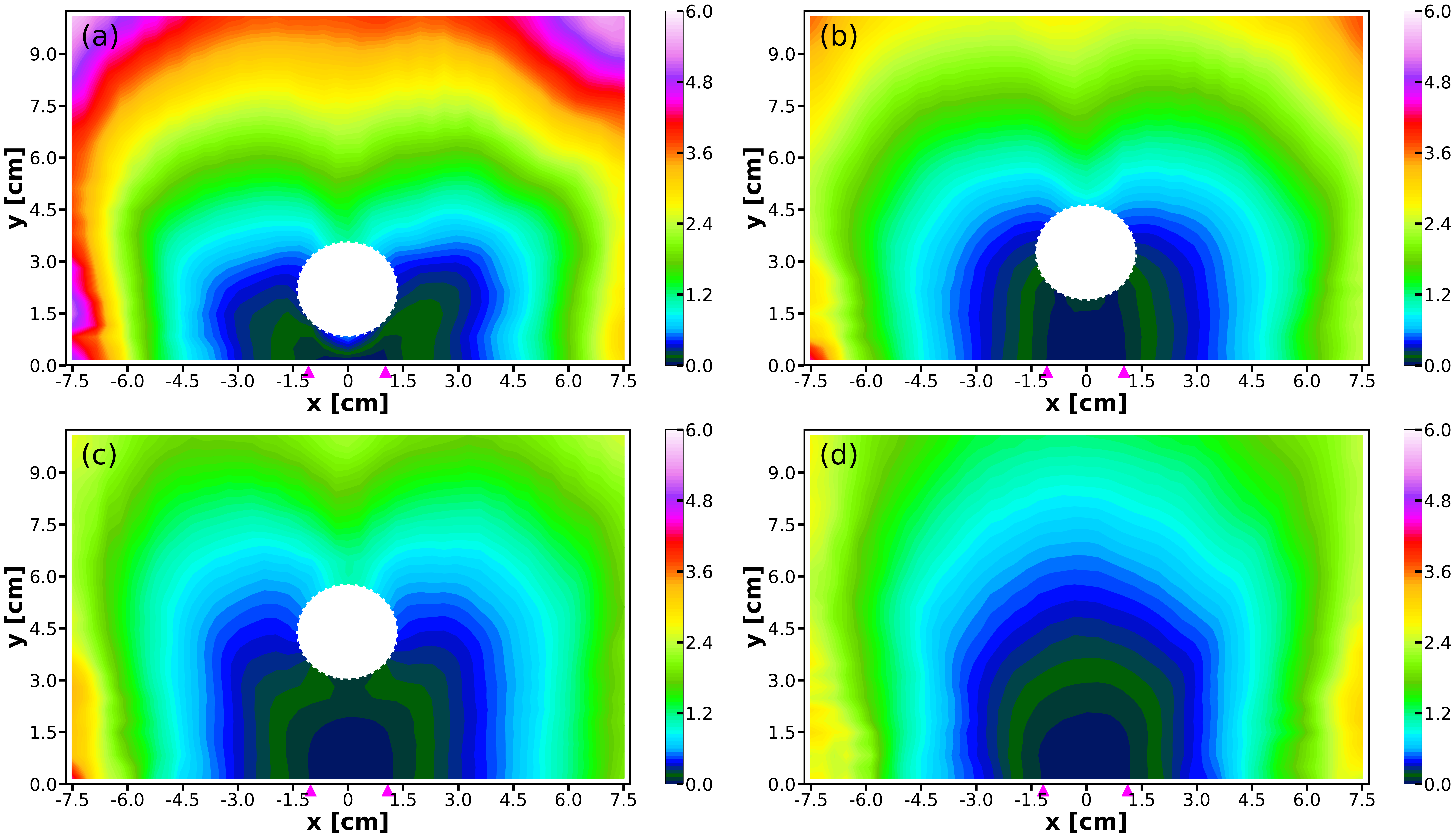}
\caption{\label{fig:esc_21} Distribution of escape times of HGS in the silo of $W=21$~mm orifice width with an obstacle at heights of (a)  $H=17.5$~mm, (b) 27.5~mm, and (c) 37.5~mm. (d) shows the escape times when the obstacle is absent. The escape times are averaged over the period where the silo was emptied from one half to one quarter. The color bar gives the time in seconds.}
\end{figure*}

In all cases, the escape times of particles starting behind the obstacle are clearly retarded with respect to particles at the same height that start above the edges of the obstacle. This is intuitively clear. One notes, however, that this retardation is much more pronounced for the
hard grains (Fig.~\ref{fig:esc_h5}(c)) than for the HGS.
Figures~\ref{fig:esc_14} and \ref{fig:esc_21} visualize the influence of the obstacle position. 
Compared to the situation without obstacle shown in subfigures (d), the obstacle
in the highest position (c) supports a somewhat faster escape of the particles at the two sides, by blocking the central downflow.
It thus leads to smaller escape time gradients along the width of the silo. One can speculate that this effect of smoothing out the distribution of the escape times by a distant obstacle could be beneficial for the escape dynamics of the animate objects: Even if the average escape time does not decrease in presence of the obstacle, there are no particular ``danger zones" from where they would not have a chance to escape in time, even if their initial distance to the exit is similar to the others. Placing the obstacle closer to the outlet leads to an overall increase of the escape time, note the different color scales in Figures~\ref{fig:esc_14} and \ref{fig:esc_21} (a), (b), and (c). 

\subsection{Local packing fraction}
\label{sec:fraction}
\begin{figure}[htbp]
\centering
\includegraphics[width=0.6\textwidth]{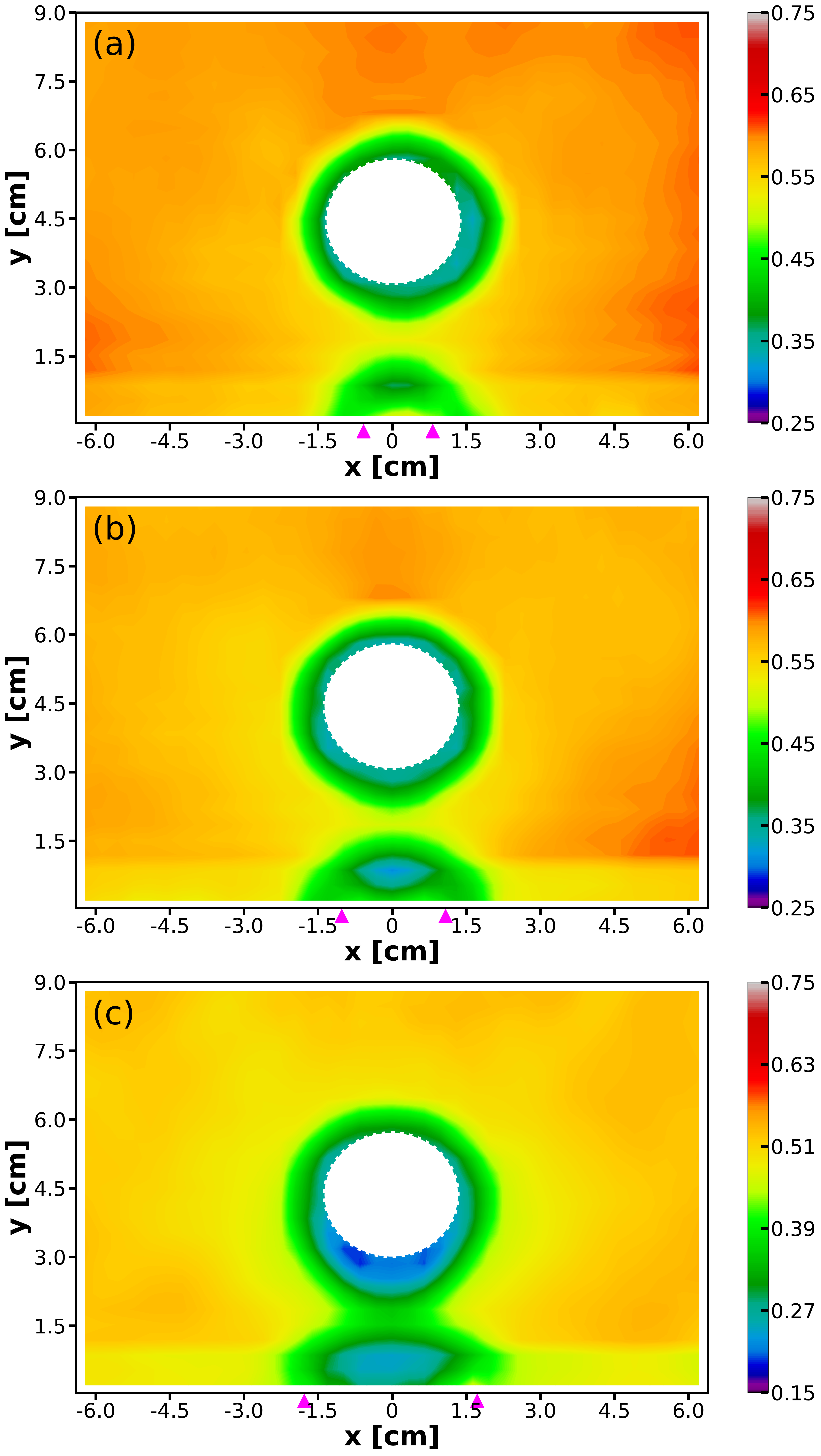}
\caption{\label{fig:dens_h5} Local 3D packing fractions of HGS for (a) $W=14$~mm and (b) $W=21$~mm.  (c) local packing fraction of ASB at $W=35$~mm. The obstacle is at height $H=37.5$~mm.
Data are time averaged over a period from half-filled to quarter-filled silo. }
\end{figure}

\begin{figure}[htbp]
\centering
\includegraphics[width=0.5\textwidth]{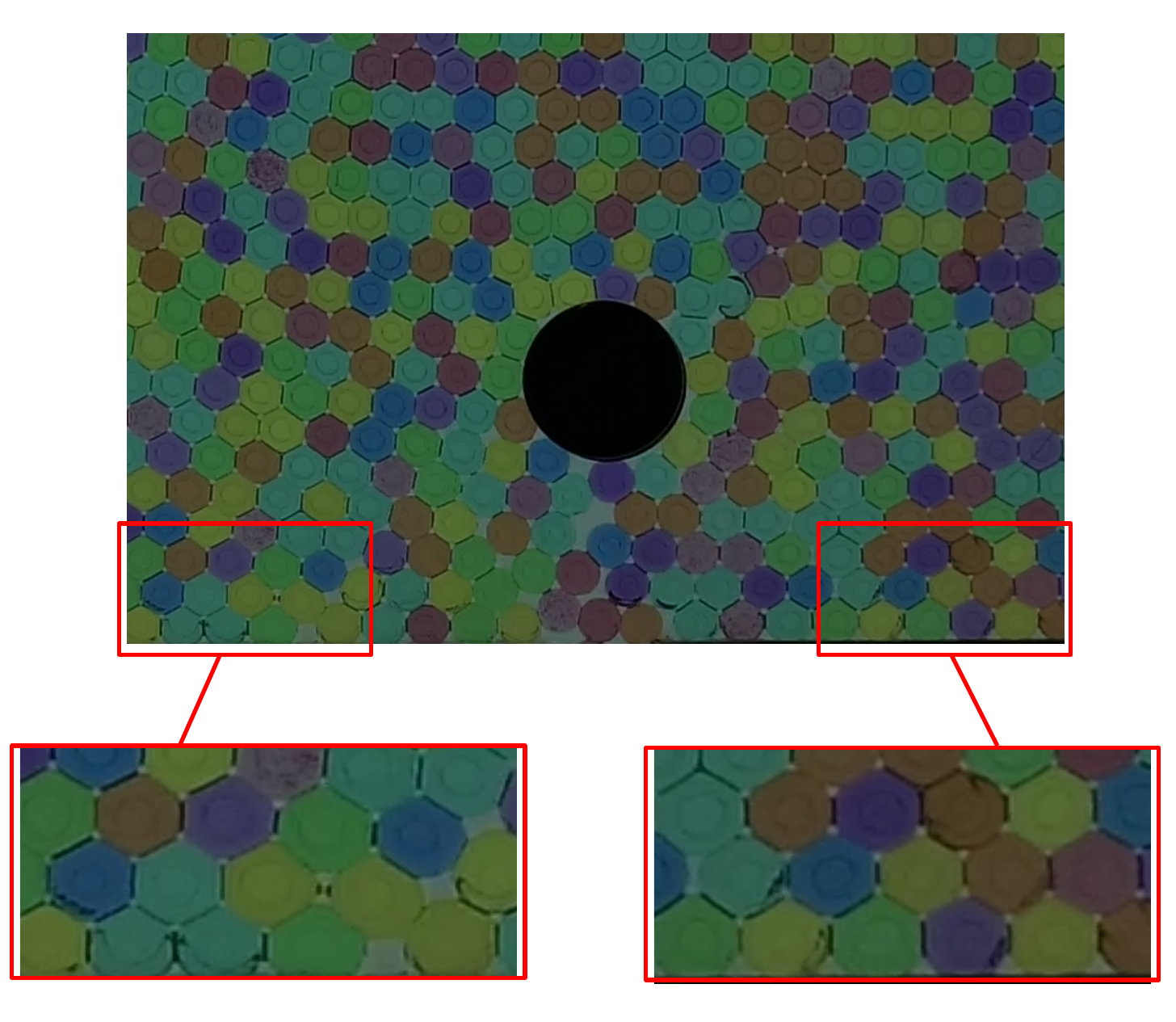}
\caption{\label{fig:zoomin} Zoomed bottom layer of HGS for $W= 14$~mm and $H= 37.5$~mm}
\end{figure}

Local packing fractions are shown in Fig. \ref{fig:dens_h5} for a fixed obstacle height of 37.5~mm. In all these plots, the packing fraction appears to be reduced in the first bottom layer. This is an artifact of the evaluation method. Indeed, the packing fraction at the borders decreases continuously towards the walls because the first layer leaves gaps at the bottom wall. Figure ~\ref{fig:zoomin} shows this in detail. Our Voronoi evaluation method averages the packing fractions over roughly one particle size.
Because the soft spheres can be deformed to a more or less hexagonal cross-section in the cell plane, their local packing can be much more efficient than for the ASB. This causes the denser packing in Fig.~\ref{fig:dens_h5}(a,b).
When one compares the packing fractions of soft and hard spheres in Fig.~\ref{fig:dens_h5}, it becomes evident that the packing immediately behind the obstacle (towards the opening), hard particles are less efficiently packed while for the soft spheres, only little difference between the packing fractions in front of and behind the obstacle (viewed in the direction of flow) is noticeable.
Towards the opening of the silo, a crescent-shaped zone of reduced packing is seen in all images.

\begin{figure*}[htbp]
\includegraphics[width=1.0\textwidth]{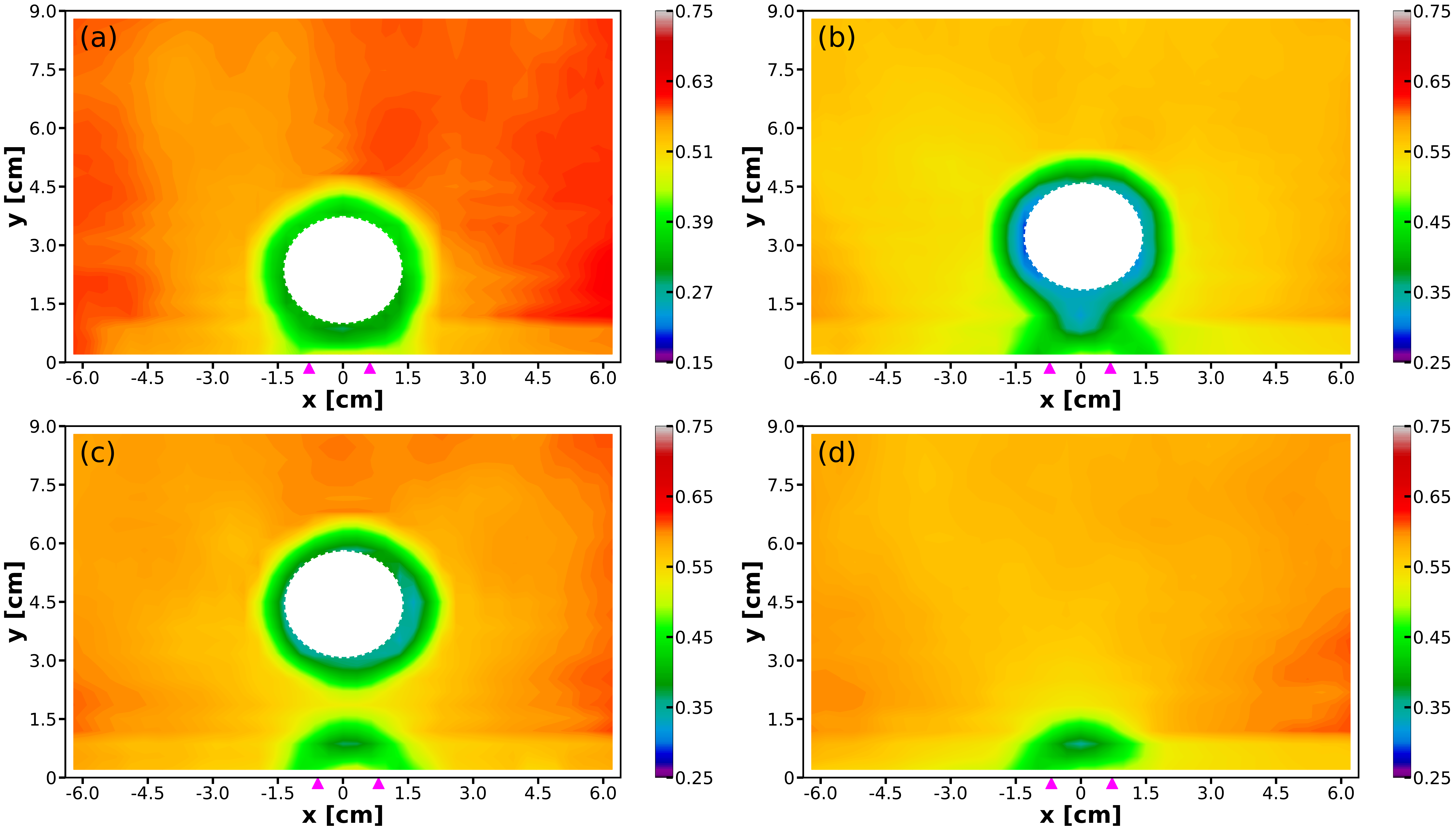}
\caption{\label{fig:dens_14}Local packing fractions of HGS in the silo of $W=14$~mm orifice with the obstacle at heights (a) 17.5~mm (b) 27.5~mm and (c) 37.5~mm, (d) shows the situation without obstacle. Packing is averaged over a discharge period from the half-filled to quarter-filled silo.}
\end{figure*}

\begin{figure*}[htbp]
\includegraphics[width=1.0\textwidth]{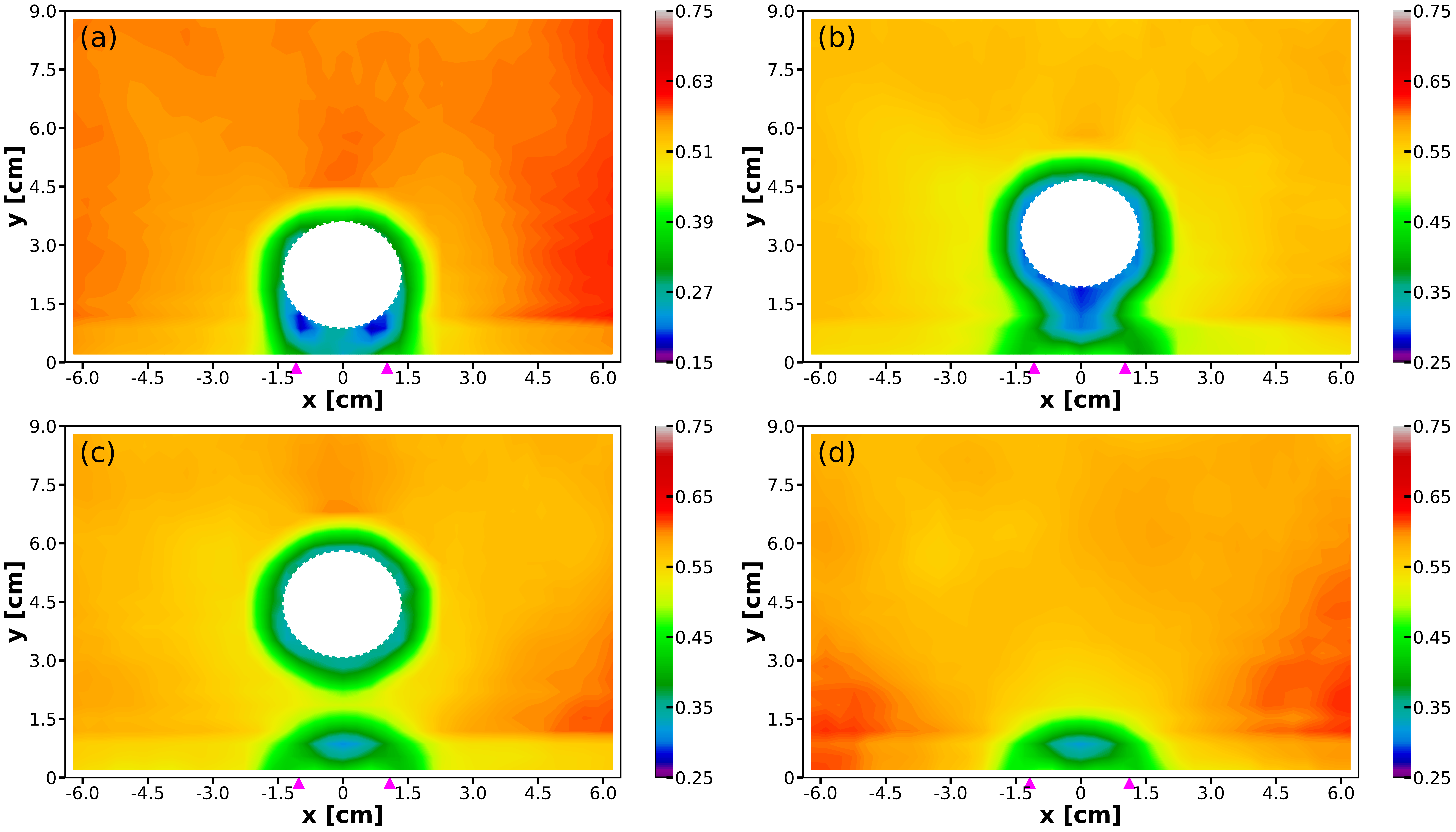}
\caption{\label{fig:dens_21}Local packing fractions of HGS in the silo of $W=21$~mm orifice with the obstacle at heights (a) 17.5~mm (b) 27.5~mm and (c) 37.5~mm, (d) shows the situation without obstacle. Packing is averaged over a discharge period from the half-filled to quarter-filled silo.}
\end{figure*}

Figures~\ref{fig:dens_14} and \ref{fig:dens_21} present the local packing fractions in the silos with 14~mm and 21~mm orifice width, respectively. 
When the obstacle is at a height of $H=37.5$~mm, there is practically no difference to the packing in absence of an obstacle in the region immediately above the outlet.
When the obstacle is placed at lower heights (b,c), it influences the packing fractions immediately above the outlet. Together with the reduced pressure above the outlet,  this leads to the lower discharge rates seen in the graphs of Fig.~\ref{fig:mass2}. 

Note that the packing fraction in the whole container increases systematically when the obstacle is placed very close to the orifice. In that case, the flow rate is reduced and there are even intermittent clogged states. During such states, the packing fraction of the
material increases \cite{harth2020intermittent}, while slow reorganization of the particle positions leads to a more efficiently packed state until the discharge continues,
even when the non-permanent clog lasts for only one or two seconds. In the time average, the packing fraction is then globally increased by a few percent.

\subsection{Kinetic stress}
\label{sec:kineticstress}
The decrease of the local packing fraction near the outlet, which is enhanced in presence of the obstacle, leads to the question whether the presence of an obstacle also induces a reduction in the magnitude of pressure not only at the silo bottom but also between the particles. This is of particular interest in view of related studies of pedestrian dynamics, where too high pressure and too strong fluctuations represent a safety hazard~\cite{Helbing2012}. With the present setup, we are unable to measure the pressure directly. Qualitatively, we observe a clear pressure reduction below the obstacle: The particle shapes are rather spherical there, indicating only little or no external stress. However, a possible increase of pressure near the gaps at the side of the outlet or even above the obstacle cannot be quantified unambiguously from our video data. Previous studies~\cite{Helbing2007,Yu2007,Garcimartin2016,garcimartin2017pedestrian} put forward the kinetic stress as a meaningful measure of the transient "crowd pressure" even in this rather dense flow, which can be directly extracted from the particle trajectories and packing fractions. Kinetic stress is a contribution to the stress tensor related to the local averaged fluctuations of velocity about their mean and the local packing fraction, and it has previously also been computed for coarse-grained systems~\cite{kirkwood1950errata,zhang2010coarse,goldhirsch2010stress}. The same concept has also been applied to flow of hard particles before~\cite{Garcimartin2016,zhang2010coarse}, in particular in silo discharge. We will apply it to the flow of our soft slippery granular material.

The kinetic stress is measured as the product of the local time-averaged velocity fluctuations and the local time averaged packing fraction, see \ref{app:kineticstress}. In our case, fluctuations of the particle velocities can reflect collisions between particles especially near the orifice, but in denser regions, they often result from dynamic rearrangements in the packing. These may originate at some distance from the actual particle position. Similar dynamics are observed in crowds~\cite{garcimartin2017pedestrian,garcimartin2017kinetic}.

\begin{figure}[htbp]
\centering
\includegraphics[width=0.6\textwidth]{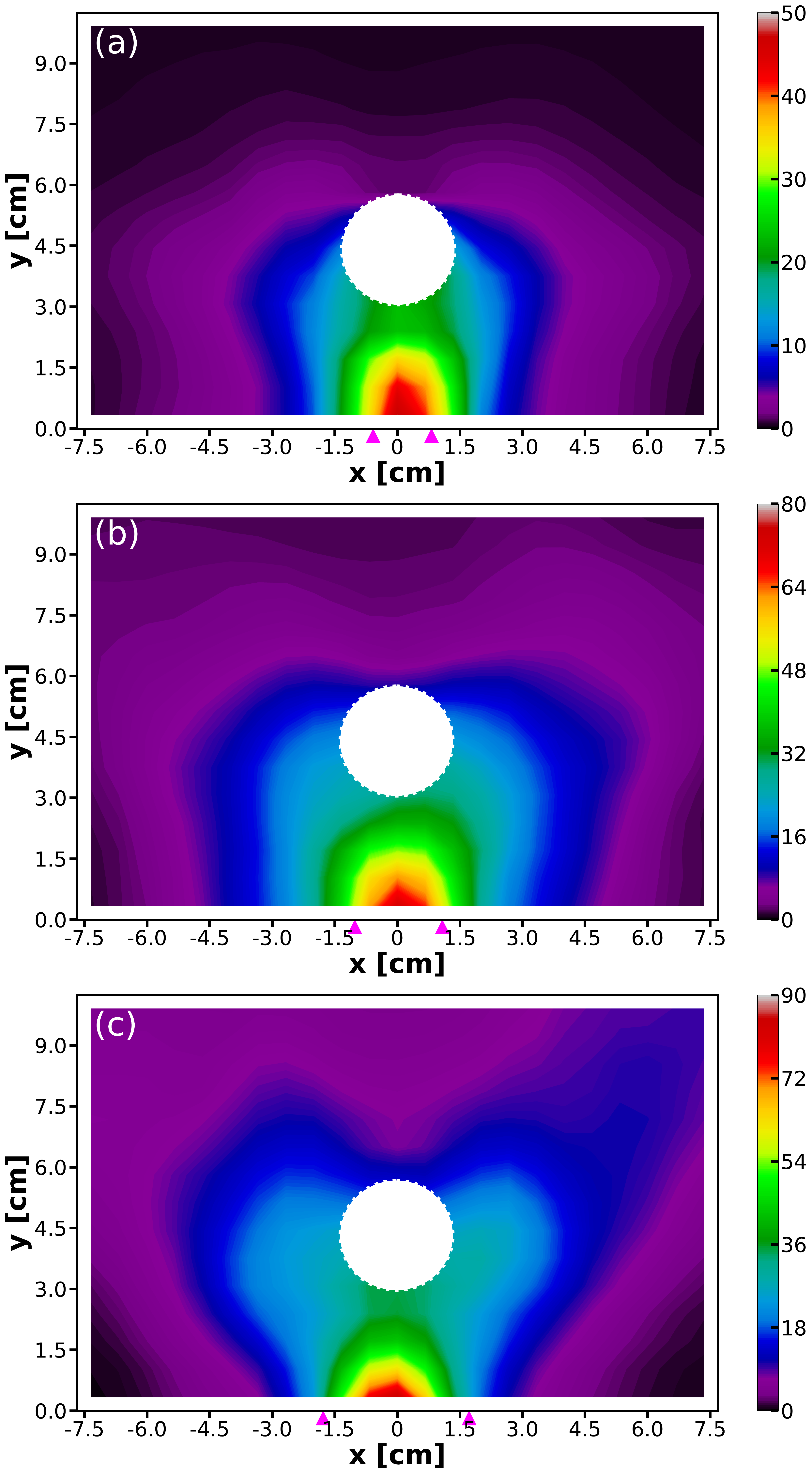}
\caption{\label{fig:kistr_h5}Kinetic stress of HGS in the silo with the obstacle at height 37.5~mm for (a) $W=14$~mm and (b) $W=21$~mm orifice width. (c) Same for ASB at $W= 35$~mm orifice wdth. Data are averaged over the discharge from one-half to one-quarter filling. Color bars give the kinetic stress in (cm/s)$^2$.}
\end{figure}

\begin{figure*}[htbp]
\includegraphics[width=1.0\textwidth]{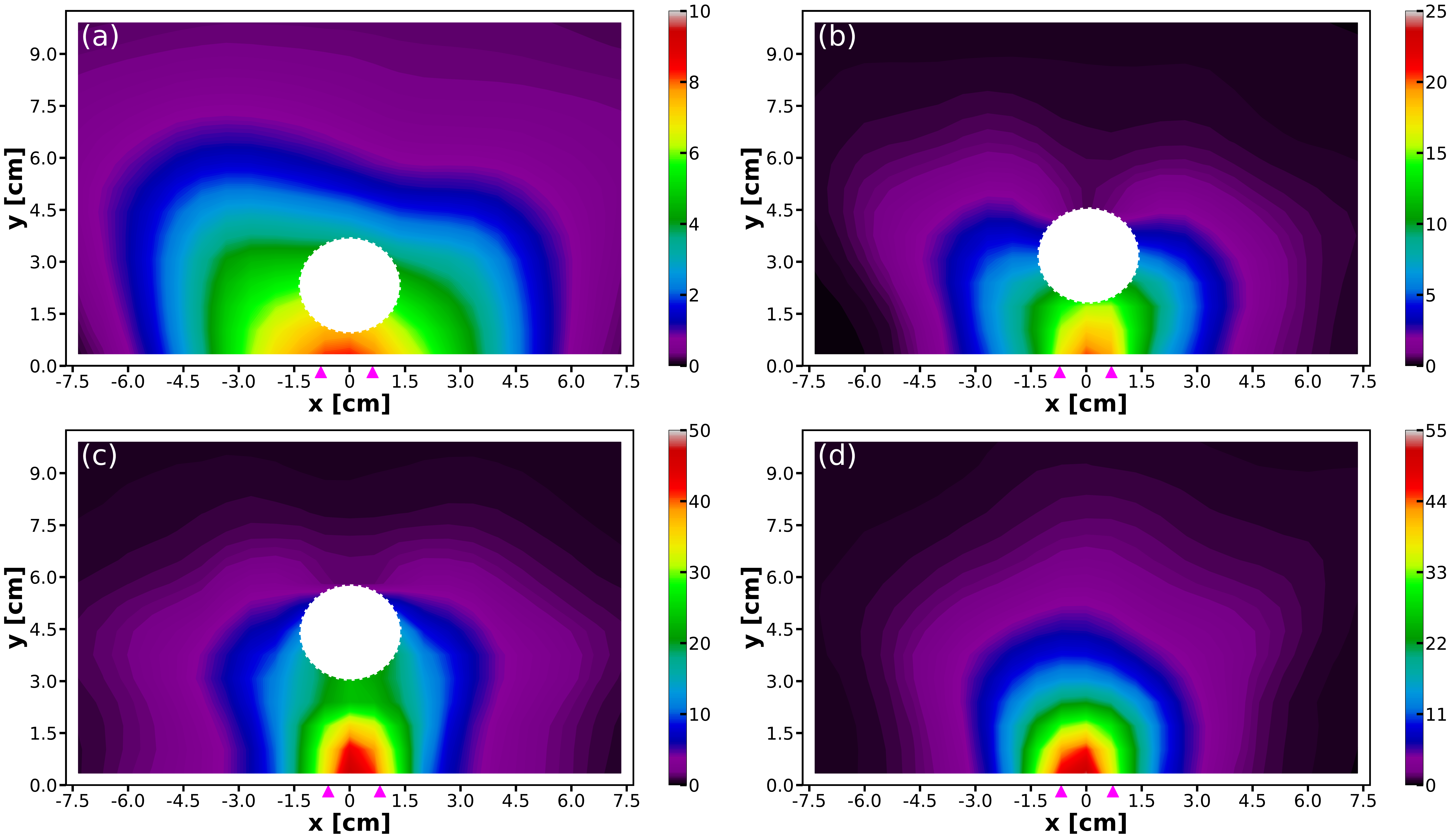}
\caption{\label{fig:kistr_14}Kinetic stress distribution of HGS in a silo with fixed $W=14$~mm orifice width and an obstacle at heights (a) $H=17.5$~mm, (b) HGS and $H=27.5$~mm, and (c) HGS and $H=37.5$~mm. (d) shows the situation without obstacle.
The kinetic stress is time averaged over the discharge from the half-filled to the quarter-filled silo, units are (cm/s)$^2$.}
\end{figure*}

\begin{figure*}[htbp]
\includegraphics[width=1.0\textwidth]{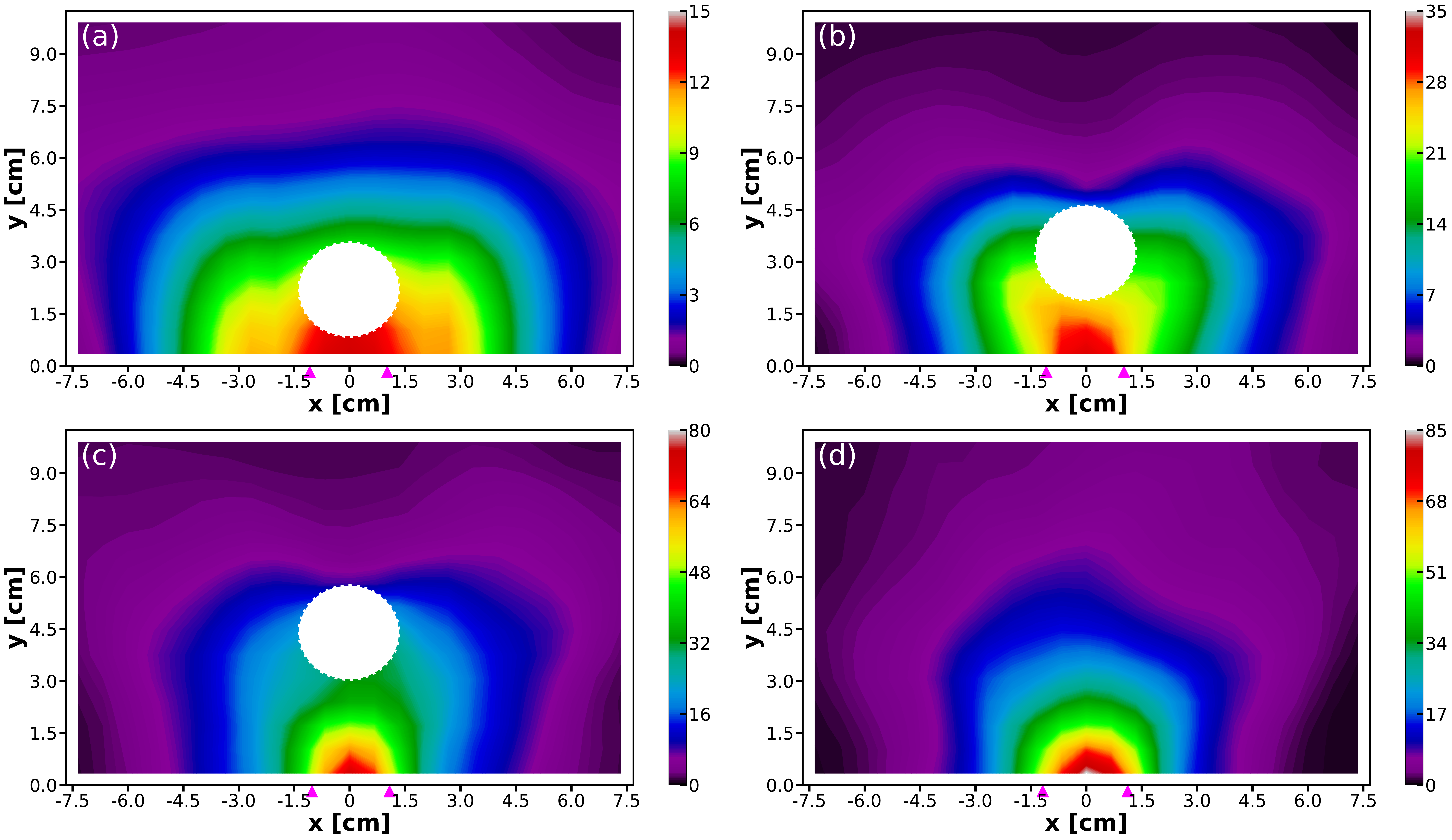}
\caption{\label{fig:kistr_21}Kinetic stress distribution of HGS in a silo with fixed $W=21$~mm orifice width and an obstacle at heights (a) $H=17.5$~mm, (b) HGS and $H=27.5$~mm, and (c) HGS and $H=37.5$~mm. (d) shows the situation without obstacle.
The kinetic stress is time averaged over the discharge from the half-filled to the quarter-filled silo, units are (cm/s)$^2$.}
\end{figure*}

In all cases analysed within this study, the kinetic stress is largest in vicinity of the orifice. Packing densities are very low here, and velocity fluctuations result from the actual outflow velocities of individual particles, which are affected both by collisions and by their dynamics upon leaving more densely packed regions. This behaviour is thus not unexpected, and the magnitude of kinetic stress increases with increasing discharge rate, i.e. when the obstacle is placed in a higher position at given orifice size (Figs.~\ref{fig:kistr_14} and~\ref{fig:kistr_21}), or when the orifice size itself is increased (Fig.~\ref{fig:kistr_h5}). The local packing fraction is lowest there (Fig.~\ref{fig:dens_h5}). Thus, the data primarily reflect the large increase of the second factor in Eq.~(\ref{eq:four}), the velocity fluctuations. 

A slight difference between the hard and soft particles is seen immediately below the obstacle in Fig.~\ref{fig:kistr_h5}, where the kinetic stress of soft particles is not as strong as for the hard grains. Another obvious difference is the lower kinetic stress in the regions towards the bottom corners, the stagnant zones, due to the lack of particle motion. The kinetic stress is lowered for both types of material directly above the obstacle where the flow is nearly stagnant.

The main influence of the obstacle is seen in Figs.~\ref{fig:kistr_14} and \ref{fig:kistr_21}, which show the situation for different obstacle heights. When the obstacle position is lowered, the kinetic stress maximum above the orifice 
splits into two regions in front of the two gaps between obstacle and the edges of the orifice. 
The gradient of the kinetic stress distribution resembles that of the distribution of escape times but is pointing in the opposite direction. Again, the pronounced low kinetic stress regions of hard grains in the bottom corners are a consequence of the stagnant zones. 

\section{Conclusions}
\label{sec:conclusions}
The distributions of velocities, escape times, granular packing fractions and kinetic stress during silo discharge have been analysed in this study. The focus of the experiments has been laid upon the effects of an obstacle in front of the outlet on the discharge of elastically deformable grains. 

When the obstacle is close enough to the outlet, it reduces the discharge rate of the particles. This is similar to previous observations of hard particle discharge from a quasi-2D silo \cite{endo2017statistical}. In contrast to reports on pedestrian evacuation or some discharge experiments with hard particles, we have not found any configurations where clogging was suppressed by the obstacle. In contrast, the obstacle can induce clogging in geometries when it would not occur in absence of the barrier. 
Interestingly, two effects can lead to clogging when the obstacle is present. First, when the gaps between the obstacle borders and the edges of the orifice become smaller than the orifice width, clogs can form in both gaps, blocking the outflow completely. The peculiarity of this situation is that when only one of the gaps is blocked, this clog may resolve when the outflow proceeds through the opposite gap. This persistent flow changes the positions of the particles above the obstacle and may break the balance of the
material forming the clog. Thus, one often 
finds long-persistent but non-permanent clogs (lasting for several seconds, even more than 30 seconds). Since we were not interested in the distinction of permanent clogs and long-lasting congestion, the experiments were stopped when a completely clogged state lasted for several seconds. This does not exclude that some of these clogged states dissolved after a longer 
period, as a result of some creeping motion in the silo.

A second scenario was discovered even when the gap between obstacle and outlet edge was considerably larger than the orifice width $W$, and the latter being two particle diameters ($\rho=2$). For such a ratio of orifice width and particle diameter, clogs are not expected in absence of an obstacle \cite{ashour2017silo}. With the obstacle, the soft particles pass the two lateral gaps without clogging, but clogs are found to occur at the orifice. The obstacle screens the particles below, at least partially, from the pressure of the overlying material. This is clearly seen in the images, where the HGS below the obstacle appear nearly circular while they are squeezed to nearly hexagonal cross sections in the regions at both sides (see, e.g., Fig.~\ref{fig:clogs}(c)). Since it is well known that the discharge of soft grains through narrow orifices is pressure dependent, it is intuitively clear that the outflow is blocked below a sufficiently low-positioned obstacle. For hard grains, where the clogging is not pressure dependent, such an effect can be excluded.

\section{Outlook}
\label{sec:Outlook}
This study discussed principal effects of an obstacle on velocities, escape times, granular packing fractions and kinetic stresses in a quasi-2D silo filled with soft spheres. 
Here, we have focused on the general features and demonstrated, what consequences can be expected. With respect to the 'faster-is-slower' effect observed, e.~g., in pedestrian dynamics  \cite{garcimartin2014experimental}, we have not found any such signs in the soft spheres system. Rather, the opposite is found. A reduction of the pressure above the orifice retards the discharge.

Interestingly, one minor effect was observed for the obstacle placed further from the orifice, viz. the decrease of the escape time for the regions near the bottom silo walls and overall smoothing of escape time distribution profile. This could be potentially beneficial for the problem of the evacuation of living objects by eliminating the ``danger zones'' from where timely escape would not be possible.

An increasing discharge rate by properly placed obstacles has been found for hard particles \cite{alonso2012bottlenecks} in special geometries. The reason for this counter-intuitive behavior of rigid materials is the reduced packing fraction near the orifice below the obstacle. 
From our observations, we conclude that a similar effect will not be found with elastic spheres, because the reduced packing fraction and pressure below the obstacle reduce the discharge rate in any case. An experimental confirmation with the geometry used in Ref.~\cite{alonso2012bottlenecks} is yet to be performed. The same applies to a more detailed characterization of escape times 
and the so-called faster-is-slower effect \cite{garcimartin2017kinetic,oh2017faster}. 

A systematic study of obstacle geometries is in general desirable. One extension of the present work could be investigations of the flow around obstacles placed far above the outlet, so that the surrounding flow field resembles plug flow. Moreover, different obstacle sizes and obstacle shapes \cite{endo2017shapes} need to be explored.

\section*{Acknowledgements}
 
This project received funding from the European Union's Horizon 2020 research and innovation program under the Marie Skłodowska-Curie grant agreement No. 812638, CALIPER. 
D. P. acknowledges funding by DLR within project EVA (50WM2048). 
The authors cordially thank Torsten Trittel for important contributions to the construction of the setup. The content of this paper reflects only the authors' view and the Union is not liable for any use that may be made of the information contained therein. 
\vspace{5mm}

\noindent\includegraphics[width=0.09\textwidth]{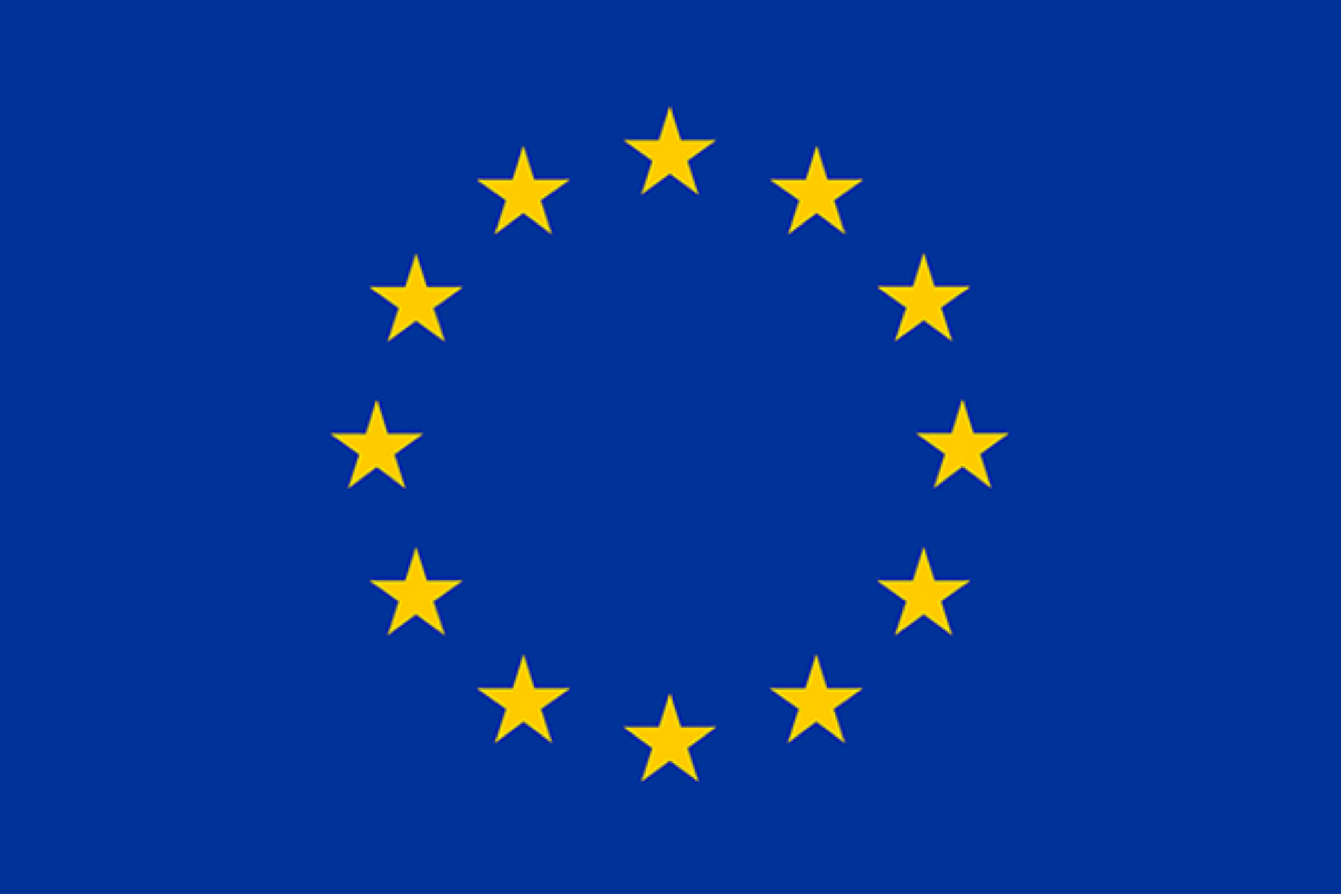}

\appendix

\section{Data analysis}
\label{app:DataAnalysis}

\subsection{Machine-Learning aided image segmentation}
\label{app:Mask-rcnn}
The optimal way to extract the information of the outflow dynamics is to perform an instance segmentation of consecutive images of the particles in silo, i.e. to find the shape the image region occupied by each particle. The positions of particles (their respective region centroids) can then be tracked in time and assembled into trajectories, provided that the frame rate is sufficiently high. Furthermore, the particle shapes can in principle provide additional information, e.g., on the pressure field
\cite{Walker2015contacts}. 

For the instance segmentation, the MASK R-CNN artificial neural network was employed. This type of network is used in multiple applications where  detection accuracy is required, especially for  particles with complex shapes \cite{Bischke2019,Qin2021mask}. We have recently applied MASK R-CNN to track the particle positions and orientations in a 3D gas of elongated particles \cite{Puzyrev2020RCNN}. To build the data set, around 30 frames with 300 to 400 particles each were manually annotated. Then, the MASK R-CNN pre-trained on the MS COCO \cite{MS_COCO} image data set was trained for around 100 epochs on the augmented images. 

\begin{figure*}[htbp]
\includegraphics[width=1.0\textwidth]{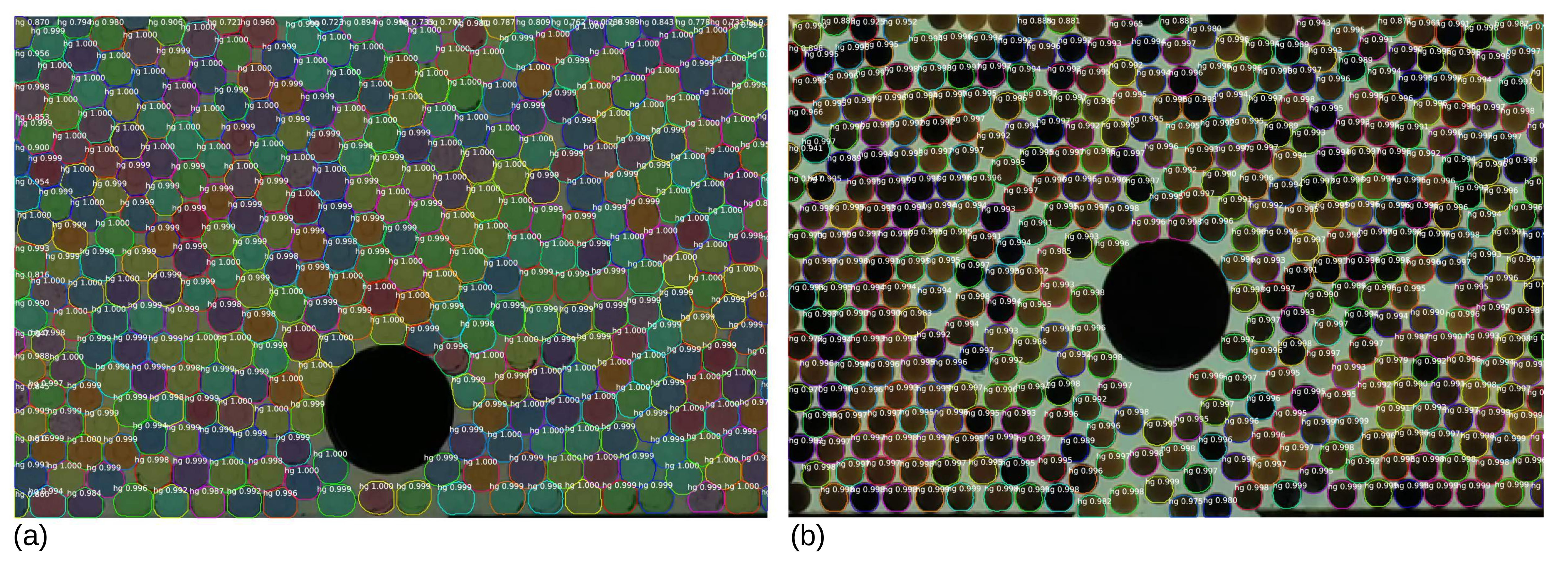}
\caption{\label{fig:detect}Examples of automatic particle detection by our MASK R-CNN trained model: Detection of (a) all soft particles, and (b) all hard particles in typical frames, by the same network.}
\end{figure*}

As expected, even with a relatively small set of training data, MASK R-CNN excels in the detection of both hard particles of round shape and soft particles with deformed shapes, see Fig.~\ref{fig:detect}. This is due to the relative simplicity of the frames, where clear borders between the objects are present. The background is clean and the image contains only a low amount of visual noise.

\subsection{Velocity fields}
\label{app:velocities}
After all particles in each frame have been detected by a trained MASK R-CNN network, we use the {\em Python trackpy} package \cite{Trackpy} to assign labels to all particles and to track them. From the trajectories of particles appearing in the image sequences, the displacement of each particle between successive frames can be determined. 
The time interval between the frames is the reciprocal $1/f$ of the frame rate $f$. 
The dynamics of particle $i$ with velocity $\vec{v}_i(t)$ at position $\vec{x}_i(t)$ is given by the equation of motion
\begin{equation}
\vec{v}_i(t) = \frac{d\vec{x}_i(t)}{dt}
\label{eq:two}
\end{equation}
In principle, it is always possible to increase the accuracy of 
velocities of slow particles with small displacements $dx_i$ by evaluating every second or third frame etc. 

\subsection{Escape time}
\label{app:escape}

The escape time map measures the expected time for a particle initially positioned in a location $(x,y)$ to reach the outlet. Depending on the presence and positioning of an obstacle, it provides a direct measure how rapidly the exit is reached. 
We again restrict ourselves to the period between half and quarter filling of the container. All particles have already been detected and tracked in the previous steps of evaluation. However, we distinguish 2 types of grains: Those, which reach the outlet in the evaluation period and those which do not. For the first, it is straightforward to build an escape time map, because their initial position and the time until the particle exits the container can be directly measured. However, if only these particles would be considered, the number of particles contributing to an average escape time decreases with increasing distance to the outlet, increasing the statistical error. Thus, we adapt a two-step procedure to calculate the full escape time map.

First, we augment the particle trajectories with additional information: For each point in the trajectory, we add the time until the end of the trajectory, the distance to the mid-point of the outlet, distance to the endpoint of the trajectory (within the selected measurement period) and the time until the last point of the trajectory. Then, we sort out all particles which reach the outlet. Here, the escape time for each point of the trajectory is immediately known. The other particles, we calculate an expected evacuation time based on the statistics of the previous data. In order to do that, we define a small range of distances from the orifice and find all trajectories ending within this range. The expected escape time along each of these trajectories is determined as follows: The end-point $(x_e,y_e)$ of the trajectory $i$ is determined, and all previously considered trajectories are searched for a point which comes into close vicinity of this endpoint. Then, the escape times from those previous trajectories to the exit are averaged, giving a value $t_e(x_e,y_e)$. To obtain the escape time for each point on trajectory $i$, we add $t_e$ to the times until the end of the trajectory for all points. This procedure is repeated with all trajectories ending in the search range.  Next, the distance range of trajectory endpoints is increased and the procedure is repeated. In this way, an expected escape time is constructed for all points in the evaluation area.

\begin{figure*}[htbp]
\begin{adjustbox}{max width=\textwidth}
\begin{tikzpicture}[node distance=1.5cm,
    every node/.style={fill=white, font=\sffamily}, align=center]
  \node (start)             [activityStarts]              {list of particle information};
  \node (onCreateBlock)     [process, below of=start]          {condition for stagnant zone};
  \node (onStartBlock)      [activityStarts, below of=onCreateBlock]   {list of moving particles};
  \node (onResumeBlock)     [process, below of=onStartBlock]   {condition for particles flowing out};
  \node (activityRuns)      [activityRuns, below of=onResumeBlock]
                            {list of particles not out yet};
  \node (onPauseBlock)      [process, below of=activityRuns]
                            {search ending points close to positions of those particles flowing out};
  \node (onDecideBlock)     [process, below of=onPauseBlock] 
                            {particles got found};
  \node (onStopBlock)       [process, below of=onDecideBlock]
                            {update duration $\Delta t + \overline{\Delta t ^\prime}$};
  \node (onDestroyBlock)    [process, below of=onStopBlock] 
                            {concatenate 2 lists};
  \node (onRestartBlock)    [process, right of=onResumeBlock, xshift=5cm]
                            {list of particles flowing out};
  \node (ActivityEnds)      [startstop, left of=onCreateBlock, xshift=-5cm]
                            {list of stationary particles};
  \node (ActivityDestroyed) [startstop, below of=onDestroyBlock]
                            {list of particle information updated};     
  \draw[->]             (start) -- (onCreateBlock);
  \draw[->]     (onCreateBlock) -- node[xshift=0.5cm,text width=0.5cm]{NO}(onStartBlock);
  \draw[->]     (onCreateBlock) -- node[yshift=0.5cm,text width=0.6cm]{YES}(ActivityEnds);
  \draw[->]      (onStartBlock) -- (onResumeBlock);
  \draw[->]     (onResumeBlock) -- node[xshift=0.5cm,text width=0.5cm]{NO}(activityRuns);
  \draw[->]     (activityRuns)  -- (onPauseBlock);
  \draw[->]     (onPauseBlock)  -- (onDecideBlock);
  \draw[->]     (onDecideBlock)  -- (onStopBlock);
  \draw[->]    (onResumeBlock) -- node[yshift=0.5cm,text width=0.6cm]{YES}(onRestartBlock);
  \draw[->]    (onDestroyBlock) -- (ActivityDestroyed);
  \draw[->]    (onStopBlock.east) -| node[yshift=1.5cm,text width=3cm]
               {add updated particle} (onRestartBlock);
  \draw[->]    (onDecideBlock.west) --++(-6,0) --++ (0,3) -- node[xshift=-2.5cm, yshift=-2.4cm,text width=4cm]{expand searching range}(activityRuns.west);
  \draw[->]   (onRestartBlock.east) --++ (0.5,0) --++ (0,-7.5) -- (onDestroyBlock.east);
  \draw[->]   (ActivityEnds.west) --++ (-0.5,0) --++ (0,-10.5) -- (onDestroyBlock.west); 
\end{tikzpicture}
\end{adjustbox}
\caption{\label{fig:flowchart}flowchart for calculation of escape time}
\end{figure*}
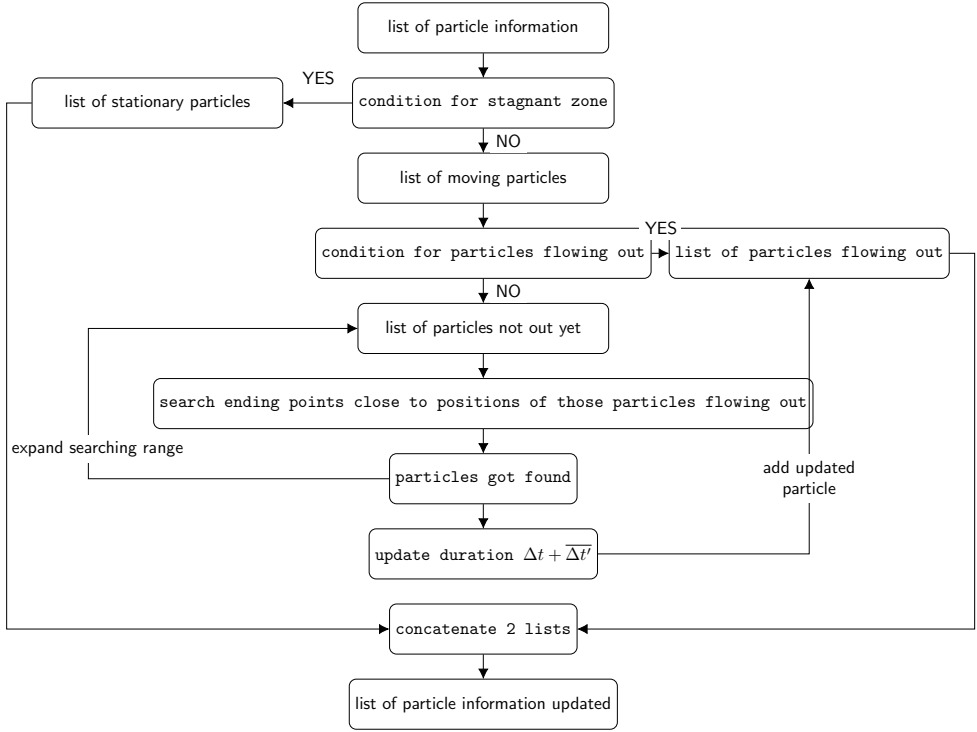

The flowchart of the calculation is shown in Fig.~\ref{fig:flowchart}.

\subsection{Local packing fraction}
\label{app:packing}
\begin{figure*}[htbp]
\includegraphics[width=1.0\textwidth]{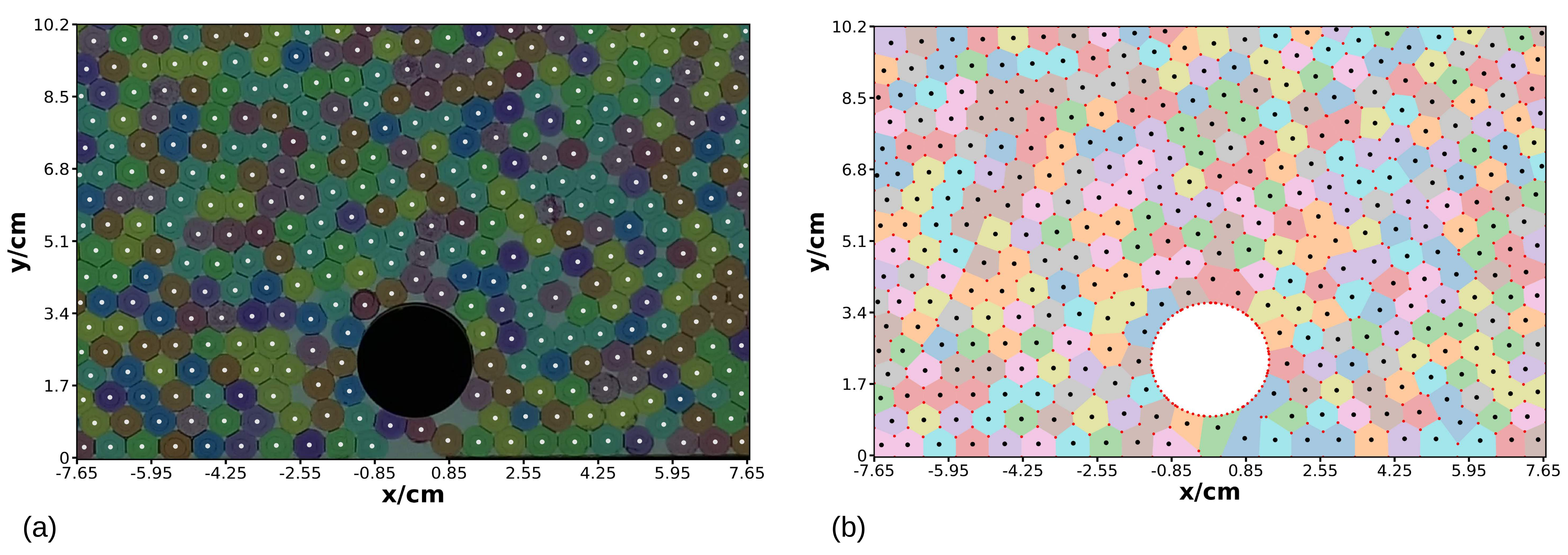}
\caption{\label{fig:voronoi}(a) location of centers of each particles in one frame randomly selected. (b) Voronoi cell for each particle in the corresponding frame.}
\end{figure*}

The method of Voronoi diagrams \cite{steffen2010methods,zhao2012spatial} has been adopted to quantify the local packing fraction of soft particles. 
In our case, after image processing, we have determined the centers of particles in each frame as shown in Fig.~\ref{fig:voronoi}(a) as seeds. Voronoi cells are calculated as shown in Fig.~\ref{fig:voronoi}(b). 
Since the depth of the silo is fixed, given by the diameter $d$ of the  particles, the Voronoi cell volume is defined as the area of a Voronoi cell $A_V$ multiplied by the silo depth $d$. The volume of each particles is $\pi d^3/6$.
Then, the local 3D packing fraction in the cell is given by  
\begin{equation}
\phi = \frac{6A_V}{\pi d^2}.
\label{eq:three}
\end{equation}

\subsection{Kinetic stress}
\label{app:kineticstress}

In our experiment, where we deal with dense flow, there is not enough information of instant stresses on each particle. Thus, traditional transient kinetic stress is not practical in our case. Instead, a time-averaged kinetic stress $\overline{\sigma_k}$ at position $\mathbf{x}$ was computed by 
\begin{equation}
\overline{\sigma_k}(\mathbf{x}) = \overline{\phi(\mathbf{x,t})} \cdot \overline {\left[  \mathbf{v}(\mathbf{x},t) - \mathbf{U}(\mathbf{x})  \right]^2} 
\label{eq:four}
\end{equation}
where the second factor represents the variance of the velocity, and
$\mathbf{U}(\mathbf{x})$ is the local time-averaged velocity. Time averaging is performed over a period where the flow can be considered quasi-stationary. Since the first factor is constant in time, averaging over several frames was simply used to minimize the statistical error.

\section*{References}

\bibliographystyle{iopart-num}
\providecommand{\noopsort}[1]{}\providecommand{\singleletter}[1]{#1}%
\providecommand{\newblock}{}

\end{document}